%% file: UVES.tex
\documentclass{aa}  

\usepackage{graphicx}
\usepackage{txfonts}

\usepackage{color,soul}

\newcommand{\Xband}{{\it X}\xspace}
\newcommand{\Vband}{{\it V}\xspace}
\newcommand{\Bband}{{\it B}\xspace}
\newcommand{\Uband}{{\it U}\xspace}

\newcommand{\snoopy}{\texttt{SNooPy}\xspace}
\newcommand{\swift}{{\em Swift}\xspace}
\newcommand{\uvot}{{UVOT}\xspace}  
 
\begin{document} 

   \title{Time-Varying Sodium Absorption in the Type Ia Supernova 2013gh}


   \author{R.~Ferretti  \inst{1}
          \and
          R.~Amanullah  \inst{1}
          \and
          A.~Goobar   \inst{1}
          \and
          J.~Johansson   \inst{2}
          \and
          P.~M.~Vreeswijk  \inst{2}
          \and
          R.~P.~Butler \inst{3}
          \and
          Y.~Cao \inst{4}
          \and
          S.~B.~Cenko \inst{5, 6}
          \and
          G.~Doran \inst{7}
          \and
          A.~V.~Filippenko \inst{8}
          \and     
          E.~Freeland \inst{9}  
          \and
          G.~Hosseinzadeh \inst{10, 11}           
          \and
          D.~A.~Howell \inst{10, 11} 
          \and
          P.~Lundqvist \inst{9}          
          \and
          S.~Mattila \inst{12, 13, 14}
          \and
          J.~Nordin \inst{15}
          \and
          P.~E.~Nugent \inst{8, 16}                           
     	 \and             
          T.~Petrushevska \inst{1}
          \and
          S.~Valenti   \inst{17}
          \and 
          S.~Vogt \inst{18}
          \and
          P.~Wozniak \inst{19}
          }

   \institute{
   	Department of Physics, The Oskar Klein Centre, Stockholm University, Albanova, SE 106 92 Stockholm, Sweden\\
	 \email{raphael.ferretti@fysik.su.se}
        \and
        Department of Particle Physics and Astrophysics, Weizmann Institute of Science, Rehovot 7610001, Israel\     
        \and
        Department of Terrestrial Magnetism, Carnegie Institute of Washington, Washington, DC 20015, USA\
        \and
        Cahill Center for Astrophysics, California Institute of Technology, Pasadena, CA 91125, USA \
        \and
        Astrophysics Science Division, NASA Goddard Space Flight Center, Mail Code 661, Greenbelt, MD 20771, USA \
	\and
	Joint Space-Science Institute, University of Maryland, College Park, MD 20742, USA\	
	\and
	Jet Propulsion Laboratory, California Institute of Technology, Pasadena, CA 91109, USA\
	\and
	Department of Astronomy, University of California, Berkeley, CA 94720-3411, USA\
	\and
	Department of Astronomy, The Oskar Klein Center, Stockholm University, Albanova, SE 10691 Stockholm, Sweden\
	\and
	Department of Physics, University of California, Santa Barbara, CA 93106-9530, USA\
	\and
	Las Cumbres Observatory Global Telescope Network, 6740 Cortona Dr., Suite 102, Goleta, CA 93117, USA\
	\and
	Tuorla Observatory, Department of Physics and Astronomy, University of Turku, V{\"a}is{\"a}l{\"a}ntie 20, FI-21500 Piikki{\"o}, Finland\
	\and 
 	Finnish Centre for Astronomy with ESO (FINCA), University of Turku, V{\"a}is{\"a}l{\"a}ntie 20, FI-21500 Piikki{\"o}, Finland\
	\and
	Institute of Astronomy, University of Cambridge, Madingley Road, Cambridge, CB3 0HA, UK\
	\and
	Humboldt-Universit{\"a}t zu Berlin, Institut f{\"u}r Physik, Newtonstrasse 15, D-12589, Berlin, Germany\
	\and
	Lawrence Berkeley National Laboratory, 1 Cyclotron Road, MS 50B-4206, Berkeley, CA 94720, USA\
	\and
        Department of Physics, University of California, One Shields Avenue, Davis, CA 95616, USA\
	\and
	UCO/Lick Observatory, Department of Astronomy and Astrophysics, University of California, Santa Cruz, CA 95064, USA\
	\and
	Los Alamos National Laboratory, MS D436, Los Alamos, NM 87545, USA\
             }

   \date{Received February 19, 2016; accepted May 4, 2016}

  \abstract
   {Temporal variability of narrow absorption lines in high-resolution spectra of Type Ia supernovae (SNe~Ia) 
   is studied to search for circumstellar matter.
   Time series which resolve the profiles of absorption lines such as Na~I~D or Ca~II~H\&K
   are expected to reveal variations due to photoionisation and subsequent recombination of the gases.
   The presence, composition, and geometry of circumstellar matter may hint at the elusive progenitor system 
   of SNe~Ia and could also affect the observed reddening law.
   }
   {To date, there are few known cases of time-varying Na~I~D absorption in SNe~Ia, all of which occurred 
   during relatively late phases of the supernova evolution. 
   Photoionisation, however, is predicted to occur during the early phases of SNe~Ia, when the supernova peaks in the ultraviolet.
   We therefore attempt to observe early-time absorption-line variations by obtaining high-resolution spectra of SNe 
   before maximum light.
   }
   {We have obtained photometry and high-resolution spectroscopy of 
   SNe~Ia 2013gh and iPTF~13dge, to search for absorption-line variations.
   Furthermore, we study interstellar absorption features in relation to the observed photometric colours 
   of the SNe.
   }
   {Both SNe display deep Na~I~D and Ca~II~H\&K absorption features.
   Furthermore, small but significant variations are detected in a feature of the Na~I~D profile of SN~2013gh.
   The variations are consistent with either geometric effects of rapidly moving or patchy gas clouds 
   or photoionisation of Na~I gas at $R\approx10^{19}$ cm from the explosion.
   }
   {Our analysis indicates that it is necessary to focus on early phases to detect 
   photoionisation effects of gases in the circumstellar medium of SNe~Ia.
   Different absorbers such as Na~I and Ca~II can be used to probe for matter at different distances from the 
   SNe. The nondetection of variations during early phases makes it possible to put limits 
   on the abundance of the species at those distances.
   }

   \keywords{
     supernovae: general --
     supernovae: individual: SN~2013gh, iPTF~13dge --        
     circumstellar matter --
     dust, extinction
               }

   \maketitle
%

\section{Introduction}

High-resolution spectra can be used to probe the material along 
the line of sight to supernovae (SNe).
The properties of the interstellar medium (ISM) of the host galaxies and also the environment of the 
SNe themselves can be studied this way.
Many high-resolution spectra of Type Ia supernovae (SNe~Ia) reveal common absorption lines of gases in the cold ISM.
For example, the Na~I~D doublet, Ca~II~H\&K, as well as 
a number of diffuse interstellar bands (DIBs) are frequently detected.
Since matter in the vicinity of SNe~Ia will be exposed to intense ultraviolet (UV) flux 
during the early phases of the explosions,
elements such as Na~I are expected to be ionised.
Thus, variations in the profiles of Na~I~D detected in multi-epoch high-resolution spectra of 
SNe~Ia have been interpreted as photoionisation of and later recombination to Na~I gas
in the circumstellar (CS) environment. 

The first detected example of varying Na~I~D lines along the line of sight of a SN~Ia was in relatively late-time high-resolution spectra of 
SN~2006X \citep{2007Sci...317..924P}. 
On a timescale of weeks after peak brightness, several Na~I components significantly increased in column density.
The unusual light curve \citep[as discussed by, e.g.,][]{2011ApJ...735...20A} and light echoes 
\citep{2008ApJ...677.1060W} from this supernova (SN) have strengthened the detection of material in the vicinity of the explosion.
Geometric effects were initially ruled out as an explanation for the variations owing to the absence 
of corresponding Ca~II~H\&K features changing with time.
However, it has been pointed out that the geometry of clouds with depleted Ca~II 
could cause the variations observed in Na~I~D \citep{2008AstL...34..389C}.

SNe~1999cl \citep{2009ApJ...693..207B} and 2007le \citep{2009ApJ...702.1157S} 
are further examples 
in which varying Na~I~D absorption have been detected. 
PTF~11kx, albeit a peculiar SN~Ia, also showed significant variations in absorption lines \citep{2012Sci...337..942D}.
More recently, SN~2014J exploded in the unusually dusty environment of 
M82 \citep{2014ApJ...784L..12G,2015ApJ...805...74B}. 
The SN light allowed for detailed studies of the ISM composition along the line of sight to M82 \citep{2015ApJ...799..197R}  and 
its peculiar extinction law \citep[e.g.,][]{2014ApJ...788L..21A,2014MNRAS.443.2887F}.
High-resolution spectra have revealed a K~I feature with decreasing depth \citep{2015ApJ...801..136G},
while the corresponding Na~I~D feature appeared not to change.
Although this may point to photoionisation of K~I atoms at $\sim10^{19}$ cm from the explosion 
and light echoes may have been observed at similar distances \citep{2015ApJ...804L..37C}, 
the analysis of late-time high-resolution spectra by \citet{2016ApJ...816...57M} strongly suggests that all identifiable
absorption features are part of extended foreground ISM clouds.

Notably, none of the above-mentioned examples of varying absorption lines resemble one another,
making a consistent interpretation of the variations difficult.
In the search for more examples of varying Na~I~D,
\citet{2014MNRAS.443.1849S} gathered multi-epoch high-resolution spectra of 14 more SNe, 
revealing no further detections. 
Thus, the few cases where varying absorption lines have been detected could simply be exceptional 
SNe occurring in unusual environments.

The rate of photoionisation can be modelled
knowing the UV flux of a SN and thus can be used to constrain 
the distance of the absorber to the explosion \citep{2009ApJ...699L..64B}. 
The subsequent recombination requires the presence of free electrons, 
which may originate from the photoionisation process or
be supplied to the CS medium by the progenitor system \citep{2009ApJ...702.1157S}. 
The presence and location of CS matter detected through absorption-line variations 
would have implications for the search for the elusive SN~Ia progenitor systems and 
also for cosmology.

Different progenitor models predict different CS environments.
For instance, the single-degenerate \citep[SD;][]{1973ApJ...186.1007W} scenario
resembles a symbiotic nova with a CS medium enriched with dust.
The environment of a double-degenerate \citep[DD;][]{1984ApJS...54..335I,1984ApJ...277..355W} 
progenitor should contain outflowing matter from the white dwarf wind \citep{2013ApJ...770L..35S},
and tidal tails can also deposit matter into the medium \citep{2013ApJ...772....1R}.
Although there is observational support for both the SD and DD models from studies of individual normal SNe~Ia
\citep[e.g.,][]{2011Natur.480..344N,2015arXiv150707261M},  the progenitor scenario of SNe~Ia is still far 
from resolved.

The light-curve shape and extinction of a SN should be affected by CS dust because photons
can be scattered multiple times.
This would give rise to steep extinction laws \citep{2008ApJ...686L.103G}, 
characterised by lower total-to-selective extinction ($R_V$) than for the same dust composition in the ISM.
Moreover, SN~Ia samples used for cosmology \citep[for a review, see][]{2011ARNPS..61..251G} 
prefer lower $R_V$ on average than what is typically observed in the Milky Way.
Low $R_V$ values can also be seen along the lines of sight to individual SNe~Ia 
\citep[e.g.,][]{2014ApJ...788L..21A,2015MNRAS.453.3300A}.  

There are observations that disfavour the presence of CS dust.
Spectropolarimetry of heavily reddened SNe~Ia 
\citep{2015A&A...577A..53P}, and the lack of infrared thermal emission 
\citep{2013MNRAS.431L..43J,2014arXiv1411.3332J} 
from some of the same events, point to the absence of dust in the CS environments.
In addition, strong limits from X-ray \citep{2014ApJ...790...52M} and radio \citep{2006ApJ...646..369P,2015arXiv151007662C} observations
point to very low mass-loss rates from the progenitor systems.
Nevertheless, there are observations suggesting the existence of outflowing matter from SNe~Ia.
A preponderance of blueshifted features in Na~I~D profiles 
of 35 high-resolution spectra \citep{2011Sci...333..856S} and 
17 intermediate-resolution spectra \citep{2013MNRAS.436..222M}
may correspond to expanding gas shells originating from the progenitor systems.
Furthermore, many SNe~Ia ($\sim25\%$) appear to have unusually high Na~I column densities,
most of which also have blueshifted Na~I~D profiles \citep{2013ApJ...779...38P}.
While reddening in the Milky Way seems to correlate well with 
Na~I column density \citep[see][]{1997A&A...318..269M,2012MNRAS.426.1465P}, 
it appears that the correlations do not hold for many SNe~Ia.
\citet{2013ApJ...779...38P} suggest that DIBs such as that at 5780~\AA\ 
are better proxies for reddening.
If the Na~I rich gas can be associated with an expanding gas shell around SNe~Ia, 
we expect to find more cases of time-varying Na~I~D owing to photoionisation.

In the following, we argue that most temporal series of high-resolution spectra of 
SNe~Ia have been taken at epochs to late to detect the photoionisation of gas 
in the CS environment of SNe~Ia.
We also present new high-resolution spectra of two SNe~Ia, of which SN~2013gh
appears to have a Na~I~D component with decreasing column density consistent with photoionisation.
The discovery and photometric properties 
of SN~2013gh and iPTF~13dge are discussed in Section~\ref{sec:disc}.
We present the high-resolution spectra in Section~\ref{sec:hires}, and
we analyse the varying Na~I~D feature in SN~2013gh.
In Section~\ref{sec:ion} we consider the possible causes for the variations, and 
we analyse the methodology of using photoionisation to detect CS matter around SNe~Ia.
Our results are summarized in Section~\ref{sec:discussion}.


\section{SN~2013gh and iPTF~13dge: Discovery and Observations}
\label{sec:disc}

\begin{figure*}
 \centering
     \includegraphics[width=8cm]{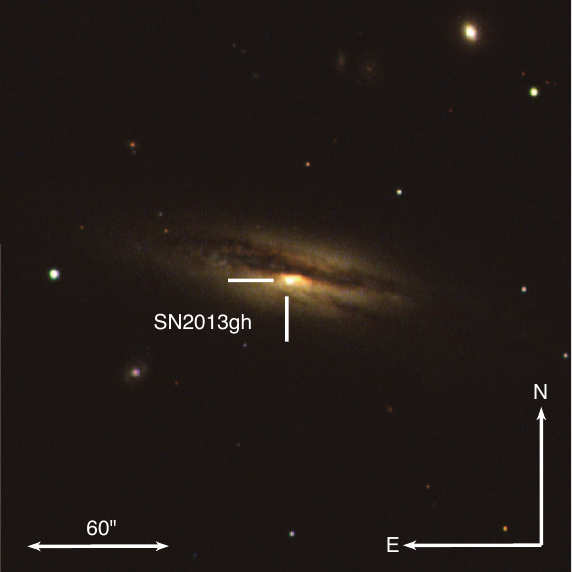}
     \includegraphics[width=8cm]{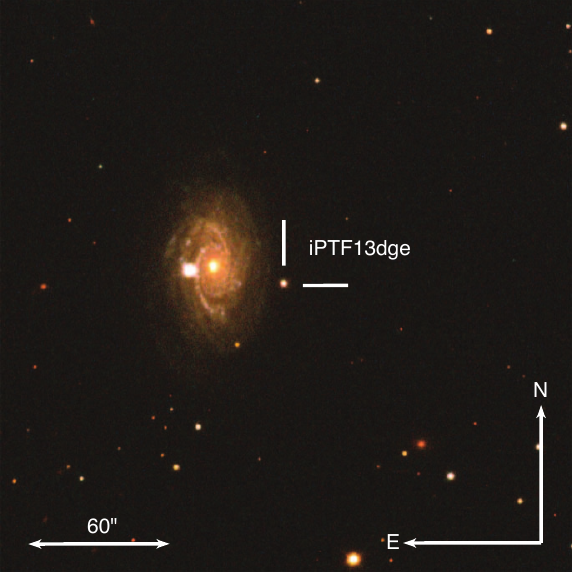}
\caption{\Uband, \Bband, and \Vband colour composite images of SN~2013gh in NGC~7183 and iPTF~13dge in NGC~1762
obtained with NOT on Sep.~10 and Sep.~23, respectively. 
       \label{fig:pics}}
 \end{figure*}

 \begin{table*}
 \centering
\begin{tabular}{l l l r l l l r r r@{}l r@{}l}
\hline\hline
\multicolumn{1}{c}{\small SN} & 
\multicolumn{1}{c}{\small $\alpha_\mathrm{SN}$} & 
\multicolumn{1}{c}{\small $\delta_\mathrm{SN}$} & 
\multicolumn{1}{c}{\small $r_\perp$} &
\multicolumn{1}{c}{\small Host galaxy} & 
\multicolumn{1}{c}{\small $\alpha_\mathrm{h}$} & 
\multicolumn{1}{c}{\small $\delta_\mathrm{h}$} &
\multicolumn{1}{c}{\small $z$} &
\multicolumn{1}{c}{\small $v_\mathrm{h}$} & 
\multicolumn{2}{c}{\small $D_{C}$} &
\multicolumn{2}{c}{\small $E(B-V)_\mathrm{MW}$} \\ 
& \multicolumn{1}{c}{\footnotesize{$(2000)$}} &
\multicolumn{1}{c}{\footnotesize{$(2000)$}} &
\multicolumn{1}{c}{\footnotesize{(kpc)}} &
& \multicolumn{1}{c}{\footnotesize{$(2000)$}} &
\multicolumn{1}{c}{\footnotesize{$(2000)$}} &
& \multicolumn{1}{c}{\footnotesize{(km s$^{-1}$)}} &
\multicolumn{2}{c}{\footnotesize{(Mpc)}} &
\multicolumn{2}{c}{\footnotesize{(mag)}}\\
\hline
{SN~2013gh} & {\small 22:02:21.84} & {\small $-18$:55:00.4} & {\small $0.6(0.1)$} & 
	{\small NGC~7183} & {\small 22:02:21.6} & {\small $-18$:54:59} & {\small 0.0088} & {\small 2635} & {\small $36$} & {\small $(2)$} &  
	\multicolumn{2}{c}{\small 0.025} \\ 
{iPTF~13dge} & {\small 05:03:35.08} & {\small $+01$:34:17.4} & {\small $9.3(0.7)$} & 
	{\small NGC~1762}      & {\small 05:03:37.0} & {\small $+01$:34:24} & {\small 0.0159} & {\small 4753} & {\small $65$} & {\small $(4)$} & 
	\multicolumn{2}{c}{\small 0.079} \\ 
\hline\hline
\multicolumn{11}{l}{}\\
\end{tabular}
\caption{%
Supernova coordinates are quoted from the discovery telegrams.  
All host-galaxy data were obtained from the NASA Extragalactic Database (NED),  
where $v_\mathrm{h}$ is the measured heliocentric recession velocity
and the Milky Way extinctions are from the \citet{2011ApJ...737..103S} calibration of the \citet{1998ApJ...500..525S} 
infrared-based dust maps.  
The host-galaxy comoving distances have been calculated based on the redshift with H$_0 = 73\pm5$~km~s$^{-1}$ Mpc$^{-1}$.
The projected distances from the host-galaxy nuclei, $r_\perp$, were calculated based on 
the host-galaxy distances and the SN offsets.
\label{tb:snsummary}}
\end{table*}

SN~2013gh \citep{2013CBET.3706....1H}, first designated as PSN~J22022184-1855004, 
was discovered by the Lick Observatory Supernova Search (LOSS) with the Katzman
Automatic Imaging Telescope \citep{2001ASPC..246..121F} in images obtained on 2013~Aug.~6.4 (UT dates are used throughout this paper). 
The SN was subsequently classified as an underluminous and/or reddened
SN~Ia one week before maximum brightness by the Las Cumbres Observatory Global Telescope (LCOGT)
with FLOYDS on Faulkes Telescope South on Aug.~11.6 \citep{2013ATel.5262....1S}. 
SN~2013gh is located in NGC~7183, 
a peculiar S0-type galaxy at redshift $z = 0.0088$ \citep{1995ApJS..100...69F}. 
The proximity to an apparent dust lane of its host galaxy suggests that the SN could be 
reddened and that ISM absorption lines may be present.  

iPTF~13dge in NGC~1762, at $z = 0.0159$ \citep{1998A&amp;AS..130..333T}, 
was discovered by the intermediate Palomar Transient Factory (iPTF) 
survey on Sep.~4.5 and classified as a SN~Ia with the 
Low Resolution Imaging Spectrometer (LRIS) on the 10~m Keck-I telescope 
on Sep.~4.6 \citep{2013ATel.5366....1C}.
The SN appears to be in the outer reaches of the host galaxy where no obvious dust 
lanes are visible. 

Here we present photometry obtained with LCOGT and with the Nordic Optical Telescope (NOT) using the 
$6.4\arcmin\times6.4\arcmin$ Andalucia Faint Object Spectrograph and Camera (ALFOSC) of both SNe as well 
as observations from the Ultraviolet/Optical Telescope 
\citep[UVOT;][]{2005SSRv..120...95R} on the \swift spacecraft \citep{2004ApJ...611.1005G} of SN~2013gh.

\begin{figure*}
 \centering
     \includegraphics[width=\textwidth]{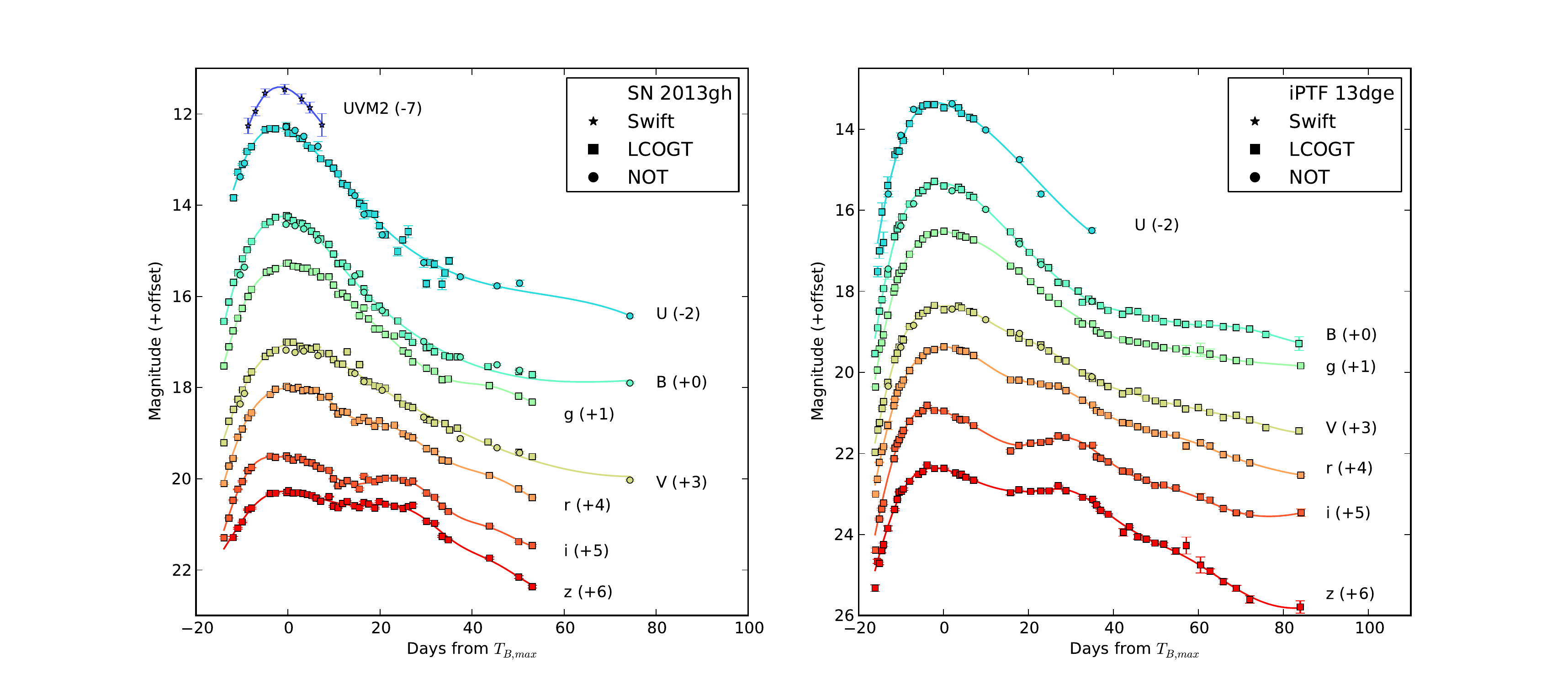} 
\caption{Light curves of SN~2013gh (left) and iPTF~13dge (right).  The data were obtained from \swift (stars), 
  LCOGT (squares), and NOT (circles).   For the {\it V}-band light curves we show the fitted splines that were used
  to obtain the colours as described in the text.  For the remaining filter we also display fitted splines to guide the eye.  
  However,  these were never used in the analysis.
       \label{fig:lcs}}
 \end{figure*}

Standard reduction with bias subtraction and flat-field correction were applied to all the ground-based data.
For the NOT data the SN flux was measured following the method outlined by \citet{2011MNRAS.416.3138V} 
using point-spread function (PSF) photometry where the host background was simultaneously estimated by fitting a polynomial.  For the LCOGT data,
on the other hand, the photometry of SN~2013gh was obtained after reference images had been subtracted, with the 
exception of the {\it U}-band data where the flux was measured directly using PSF photometry.   Similarly, the photometry for all bands of iPTF~13dge could be obtained using PSF photometry directly owing to the significant separation between the SN and the host.  
All ground-based photometry was calibrated using either Landolt standards or stellar photometry from the Sloan Digital Sky Survey.   
Furthermore, the \uvot photometry was obtained from the \swift Optical/Ultraviolet Supernova Archive \citep[SOUSA;][]{2014Ap&SS.354...89B}. The reduction is based on that of 
\citet{2009AJ....137.4517B}, including subtraction of the host-galaxy count rates, and uses the revised UV zeropoints and 
time-dependent sensitivity from \citet{2011AIPC.1358..373B}. All photometry 
is presented in Table~\ref{tab:measuredphot} and the light curves are shown in Figure~\ref{fig:lcs}.
\begin{table}
  \centering
  \input{data_table_short}
  \caption{%
    The measured photometry of SN~2013gh and iPTF~13dge.  The magnitudes are given in the natural systems of the filters and the uncertainty is $1\sigma$.
  (This table is available in its entirety in a machine-readable form in the online journal. A portion is shown here 
   for guidance regarding its form and content.)
	\label{tab:measuredphot}}
\end{table}

Having UV and optical photometry of SN~2013gh, we use the method outlined by \citet{2015MNRAS.453.3300A} to  
simultaneously fit the extinction-law parameters $R_{V}$ and $E(B-V)$ together with the time of $B$-band maximum, $t_{B_{\rm max}}$,
and the light-curve decline rate parametrised by the stretch parameter, $s$.  The fit is based on the assumption that the observed
colours of SN~2013gh can be described by the colours of the normal and unreddened SN~2011fe together with the 
\citet{1999PASP..111...63F}  extinction law.  The colours of the two objects are further compared for 
the same effective light-curve-width-corrected ``phases,'' $p$, 
which for each SN is defined as $p=(t-t_{B_{\rm max}})/(1+z)/s$, where $t$ is time in the observer's frame, $z$ the redshift of
the object, and $t_{B_{\rm max}}$ and $s$ are left as free parameters for 
SN~2013gh but kept fixed for SN~2011fe as described by \citet{2015MNRAS.453.3300A}.

The observed colours were studied for each filter with respect to the \Vband band, which could be obtained directly for the epochs
where \Vband observations were available.  For the remaining epochs the \Vband magnitudes were obtained from 
a fitted spline model to the existing data using \snoopy \citep{2011AJ....141...19B}.  
\snoopy is a software package primarily designed to fit light-curve parameters of SNe~Ia 
but can also be used to fit different models to the light-curve data.  The \Vband magnitudes that were used  to measure the 
colours are also shown in Table~\ref{tab:phot}, where column~7 indicates whether the value was obtained from data or from the 
\snoopy model.  In this table we have summarised all the data that were used to calculate the colours, and column 4 lists the measurements adopted from Table~\ref{tab:measuredphot}.  We also calculate the Galactic extinction for each point using the values from Table~\ref{tb:snsummary} as well as the cross-filter corrections, $K_{X}$, between the observed filters and the 
corresponding filter for which SN~2011fe has been observed.  The details can be found in \citet{2015MNRAS.453.3300A}. 

\begin{figure}
   \centering
      
       \resizebox{\hsize}{!}{\includegraphics{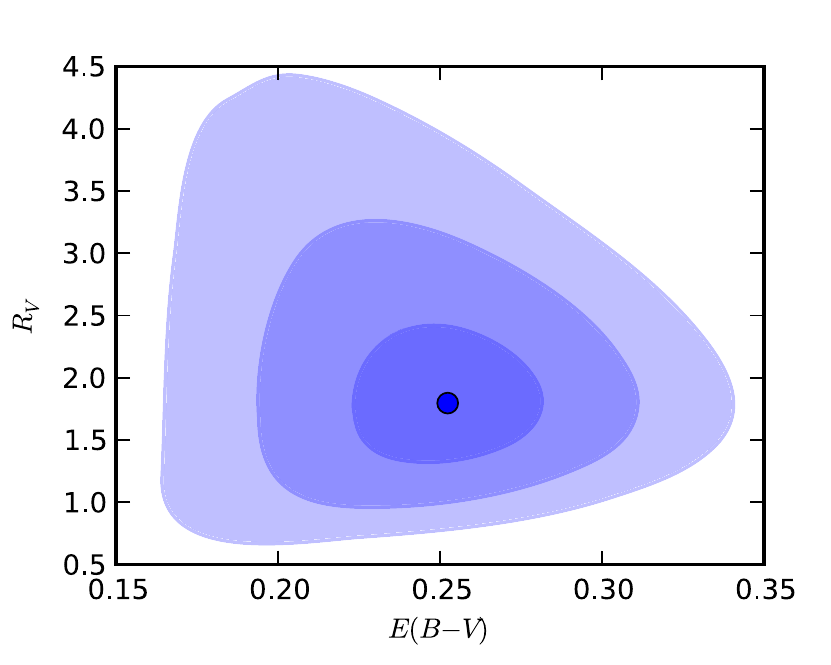}}
   \caption{The $68.3\%$, $95.5\%$, and $99.7\%$ level statistical confidence contours of the 
        $R_{V}$ and $E(B-V)$ fit to the SN~2013gh observed light-curve colours.  The point shows the best-fit value
        from Table~\ref{tab:lcs}. See text for details.}
         \label{fig:banana}
    \end{figure}

The best-fit light curve and reddening-law parameters are shown in Table~\ref{tab:lcs} using all 
data between phases $-10$ and $+35$~days from $t_{B_{\rm max}}$. In Figure~\ref{fig:banana}, the $R_{V}$ vs. $E(B-V)$ 
contours of the best fit are shown. 

iPTF~13dge, on the other hand, had very low reddening which did not allow $R_{V}$ to be well constrained.  For this SN we fit 
the reddening after fixing $R_{V} = 3.1$.   

\begin{table*}
\centering
\input{color_table_short}

  	\caption{%
  Photometry of SN~2013gh and iPTF~13dge.  Here, column 2 shows the effective light-curve-width-corrected phase.
  In column~4, \Xband is the natural magnitude in the filter specified in column~3, while columns 5--6
  ($A^\mathrm{MW}_\mathrm{\Xband}$ and $K_\mathrm{\Xband}$) are the Galactic absorption and the filter correction ($S$-correction) 
  to the corresponding rest-frame filter for SN~2011fe as described by \citet{2015MNRAS.453.3300A}.  The corrected magnitude 
  can be obtained as $\mathrm{\Xband} - A^\mathrm{MW}_\mathrm{\Xband} - K_\mathrm{\Xband}$.
  All corrections have been calculated after the SN~2011fe template has been reddened with the best-fit \citet{1999PASP..111...63F} law, for each SN.  Columns 8--10 show the \Vband magnitude and corrections for the same phase.  
  Column~7 specifies whether the \Vband magnitude was measured for the same epoch (D) or if it was calculated using the 
  \snoopy model (M).  The \Vband magnitude is only shown for data points used in the colour analysis, with phases 
  between $-10$ and $+35$ days.  The corrected colour can be obtained as 
  $(\mathrm{\Xband} - A^\mathrm{MW}_\mathrm{\Xband} - K_\mathrm{\Xband}) - (\mathrm{\Vband} - A^\mathrm{MW}_\mathrm{\Vband} - K_\mathrm{\Vband})$ and can be compared with the corresponding colour of SN~2011fe shown in the
  last column in order to study the reddening laws of the SNe.
  (This table is available in its entirety in a machine-readable form in the online journal. A portion is shown here 
   for guidance regarding its form and content.)
	\label{tab:phot}}
\end{table*}

\newcommand{\mycolsep}{\hspace{0.9em}}
    \begin{table}
       	\centering
  	\begin{tabular}{l@{\mycolsep}c@{\mycolsep}c@{\mycolsep}c@{\mycolsep}c}
    	\hline\hline
	\multicolumn{1}{c}{\small SN} & {\small $t_{B_{\rm max}}$} & {\small $s$} & {\small $R_{V}$} & {\small $E(B-V)$} \\
	& {\small (MJD)} & & & {\small (mag)}  \\ 
    	\hline
	{\small SN~2013gh} & {\small 56526.9$\pm$0.2} & {\small 0.92$\pm$0.01} & {\small $1.7^{+0.6}_{-0.5}$} & {\small $0.25\pm0.03\phantom{^{\dagger}}$} \\ 	
	{\small iPTF~13dge} & {\small 56558.0$\pm$0.2} & {\small 0.96$\pm$0.01} & {\small --} & {\small $0.03\pm0.04^{\dagger}$} \\ 
	\hline\hline
	\multicolumn{3}{l}{\small $^{\dagger}$Assuming $R_{V}=3.1$.}\\
  	\end{tabular}
  	\caption{Light-curve and reddening parameters of SN~2013gh and iPTF~13dge obtained by fitting the colour curves of 
	  SN~2011fe to the observed data together with the \citet{1999PASP..111...63F} extinction model using the method 
	  outlined by \citet{2015MNRAS.453.3300A}.
	\label{tab:lcs}}
\end{table}

\section{High-Resolution Spectroscopy}
\label{sec:hires}

Three epochs of high-resolution spectra 
of SN~2013gh as well as one epoch of iPTF~13dge were obtained with the 
Ultraviolet and Visual Echelle Spectrograph  \citep[UVES;][]{2000SPIE.4008..534D} on UT2 
at the Very Large Telescope (VLT) of the European Southern Observatory (ESO).
The REFLEX (ESOREX) reduction pipeline  provided by ESO was used to reduce the raw UVES data. 
Another high-resolution spectrum of iPTF~13dge was obtained with 
the High Resolution Echelle Spectrograph \citep[HIRES;][]{1994SPIE.2198..362V}
mounted on the Keck-I telescope.
The HIRES spectrum was reduced using the XIDL reduction pipeline\footnote{\url{http://www.ucolick.org/~xavier/IDL/}}.
These high-resolution spectra along with estimated signal-to-noise ratio (S/N) and 
resolution are summarised in Table~\ref{tab:spec}.
We have used the line-modelling software 
Molecfit \citep{2015A&amp;A...576A..77S,2015A&amp;A...576A..78K} to correct for telluric features in relevant parts of the spectra. 

    \begin{table*}
       	\centering
  	\begin{tabular}{l c c l@{\,}r c r@{\,}l c c}
    	\hline\hline
    	{\small SN} & {\small Instrument} & {\small MJD} & \multicolumn{2}{c}{\small UT Date}  &
	{\small Exp. time} &   \multicolumn{2}{c}{\small Phase} & {\small $R$ $(\lambda/\delta \lambda)$} & {\small S/N}\\
	& & & \multicolumn{2}{c}{\small } & {\small (s)} & \multicolumn{2}{c}{\small (days)} & & \\
    	\hline
	{\small SN~2013gh} & {\small UVES} & {\small 56,517.3} & {\small Aug.}&{\small 13.3} & {\small $3 \times 1700$} & {\small $-8.8$}&{\small(0.2)} 
	& {\small 74,000} & {\small 56}\\ 
	{\small SN~2013gh} & {\small UVES} & {\small 56,544.1} & {\small Sep.}&{\small 9.1} & {\small $3 \times 1700$} & {\small 15.9}&{\small(0.3)} 
	& {\small 64,000} & {\small 57}\\  
	{\small SN~2013gh} & {\small UVES} & {\small 56,559.0} & {\small Sep.}&{\small 24.0} & {\small $3 \times 1700$} & {\small 29.3}&{\small (0.3)} 
	& {\small 58,000} & {\small 38} \\ 
	\hline
	{\small iPTF~13dge} & {\small UVES} &{\small 56,547.3} & {\small Sep.}&{\small 12.3} & {\small $3 \times 1700$} & {\small $-10.1$}&{\small (0.2)} 
	& {\small 69,000} & {\small 30}\\  
	{\small iPTF~13dge} & {\small HIRES} & {\small 56,551.5} & {\small Sep.}&{\small 16.5} & {\small $2 \times 600$} & {\small $-6.1$}&{\small (0.2)} 
	& {\small 60,000} & {\small 21} \\ 
	\hline\hline
	    \multicolumn{1}{l}{}\\
  	\end{tabular}
  	\caption{The obtained high-resolution spectra with stretch-corrected phases in the rest frame with respect to $t_{B_{\rm max}}$. 
	The resolution is estimated from the full width at half-maximum intensity (FWHM) of several telluric features, 
	and the S/N per pixel is measured around the wavelength of Na~I~D.
	\label{tab:spec}}
\end{table*}

Absorption features of spectra can be characterised by their equivalent widths (EWs),
which can be measured using the point-to-point formula
\begin{equation}
{\rm EW}=\sum_{i=1}^{N}\left(1-\frac{f(\lambda_{i})}{f_{c}(\lambda_{i})}\right)\Delta\lambda_{i},
\end{equation}
where $f$ and $f_{c}$ are the actual flux and continuum flux
at a given wavelength $\lambda$, respectively.
The propagated uncertainty, $\sigma_{\rm EW}$, is given by 
\begin{equation}
\sigma_{\rm EW}^{2}=\sum_{i=1}^{N}\left(\frac{\sigma_{f}^{2}\left(\lambda_{i}\right)}{f_{c}^{2}(\lambda_{i})}+\frac{f^{2}\left(\lambda_{i}\right)}{f_{c}^{4}(\lambda_{i})}\sigma_{f_{c}}^{2}\left(\lambda_{i}\right)\right)(\Delta\lambda_{i})^{2},
\end{equation}
where $\sigma_{f}$ is the 
uncertainty of the actual flux and $\sigma_{f_{c}}$ is the 
uncertainty of the continuum flux $f_{c}$.

We have found that the REFLEX pipeline produces larger uncertainties than the root-mean-square (RMS) scatter about the continua of spectra.
Over short sections of the spectra, the RMS scatter appears to be $\sim35\%$ less than the average pipeline uncertainties.
For this reason, we use the RMS scatter instead of the pipeline output uncertainties to compute EWs.

The continuum near well-defined lines such as Na~I~D and Ca~II~H\&K can be determined
by fitting a third-degree polynomial to sections bracketing the features. This method works poorly
for broad, diffuse features such as DIBs, since the features themselves are not easily separable from the continua. 
We therefore simultaneously fit a polynomial and a Gaussian function to determine the continuum with the 
polynomial parameters. 
The EWs of the DIBs can subsequently be computed using Equation~(1).
We also fit multiple Voigt profiles using VPFIT\footnote{\url{http://www.ast.cam.ac.uk/~rfc/vpfit.html}} 
to the absorption lines to determine individual components and 
column densities, where relevant.

\subsection{SN~2013gh}
\label{sec:hires_13gh}
   \begin{figure*}
   \centering
   	\resizebox{\hsize}{!}{\includegraphics{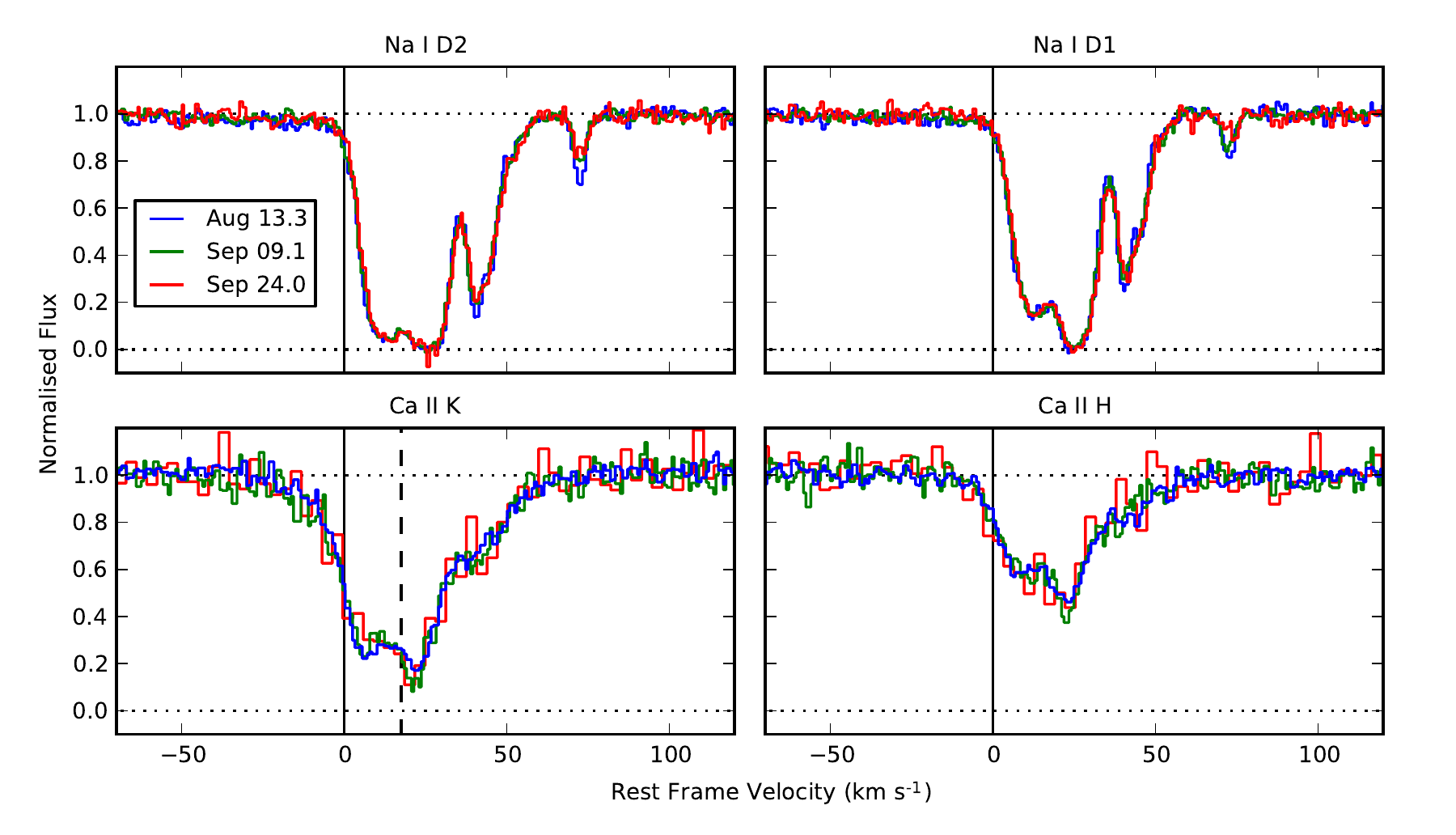}}
   \caption{The Na~I~D doublet and Ca~II~H\&K in the three UVES spectra of SN~2013gh. 
   The spectra are normalised and the Sep. 24.0 spectrum is binned for clarity in the Ca~II~H\&K subplots. 
   The solid vertical lines indicate the rest-frame wavelength of the host galaxy NGC~7183 
   \citep[according to the spectroscopic redshift in][]{1995ApJS..100...69F}.
   The dashed vertical line in the Ca~II~K subplot indicates the Milky Way rest-frame wavelength of Ca~II~H.
 }
         \label{fig:naid}
    \end{figure*}

 \begin{table}
  \centering
  \begin{tabular}{l r@{\,}l c c}
    \hline\hline
    {\small Feature}  &  \multicolumn{2}{c}{\small EW}   & {\small FWHM}  & {\small Estimated $E(B-V)$}\\
    & \multicolumn{2}{c}{\small (m\AA)} & {\small (\AA)}  & {\small (mag)} \\
    \hline
	{\small Na~I~D1} & {\small 682} & {\small $\pm$ 3} & {\small - }& {\small$>0.8^{+0.4}_{-0.3}$}\\
	{\small Na~I~D2} & {\small 816} & {\small $\pm$ 3} & {\small - } & {\small$>0.7^{+0.3}_{-0.2}$}\\
	{\small Ca~II~H} & {\small 222} & {\small $\pm$ 6} & {\small - } & {\small - } \\
	{\small DIB~5797} & {\small 26} & {\small $\pm$ 4} & {\small 0.6} & {\small$0.10^{+0.13}_{-0.07}$}\\
        {\small DIB~6196} & {\small 21} & {\small $\pm$ 7} &  {\small 0.9} & {\small - } \\
	{\small DIB~6203$^{\dagger}$} & {\small 31} & {\small $\pm$ 9} & {\small 1.8} & {\small - } \\ 
	{\small DIB~6270} & {\small 30} & {\small $\pm$ 7} & {\small 1.5} & {\small - } \\ 
	{\small DIB~6284} & {\small 66} & {\small $\pm$ 9} & {\small 2.2} & {\small - } \\
	{\small DIB~6379} & {\small 18} & {\small $\pm$ 3} & {\small 0.4} & {\small - } \\
	{\small DIB~6614} & {\small 48} & {\small $\pm$ 5} & {\small 0.8} & {\small - } \\
	{\small DIB~6661} & {\small 9} & {\small $\pm$ 3} & {\small 0.4} & {\small - } \\
    \hline\hline
    \multicolumn{5}{l}{\small $^{\dagger}$Measurement possibly blended with DIB~$\lambda$6204}.\\
  \end{tabular}
  \caption{The equivalent widths of  Na~I~D, Ca~II~H\&K, and the 
  detectable DIBs in SN~2013gh. The FWHM of each DIB has been obtained from the 
  Gaussian fit to the feature.
  The $E(B-V)$ is computed by the empirical relations by 
  \citet{2012MNRAS.426.1465P}  for Na~I~D and \citet{2015MNRAS.447..545B} for the DIB~$\lambda$5797. 
  \label{tab:ew_13gh}}
\end{table}
 
    \begin{figure}
   \centering
       \resizebox{\hsize}{!}{\includegraphics{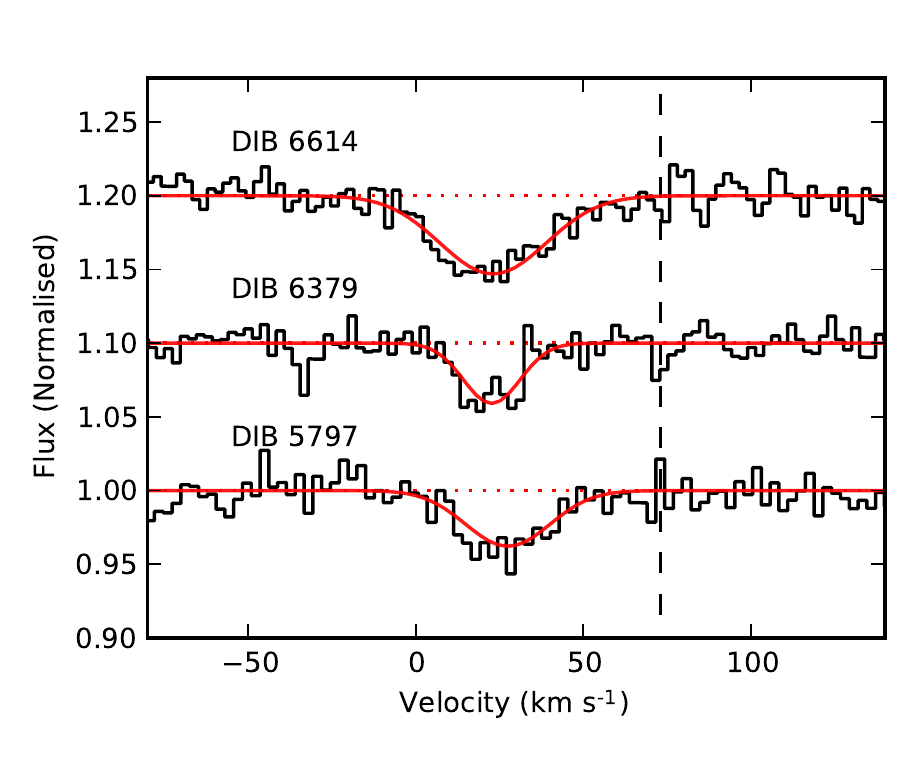}}
       \caption{Three DIBs of SN~2013gh with well-defined profiles. 
       The spectra are normalised, binned, and offset from one another. 
       The red solid lines are the best-fit Gaussian profiles of the features.
       The dashed vertical line indicates the velocity of the small feature 
       at $v\approx73$ km~s$^{-1}$ in Na~I~D.
              \label{fig:dib}}
    \end{figure}
 
   \begin{figure}
   \centering
       \resizebox{\hsize}{!}{\includegraphics{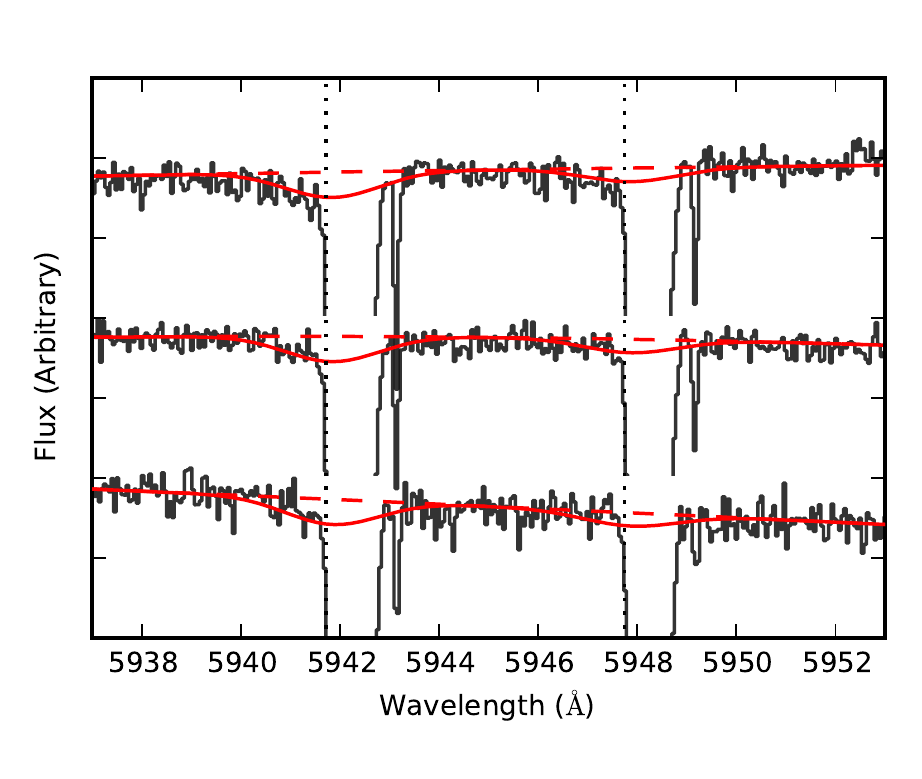}}
       \caption{
       The three spectra of SN~2013gh around the Na~I~D doublet, 
       highlighting the possible diffuse feature of the profile (red solid line). 
       The spectra have not been normalised, excluding the possibility that the broad feature is an artefact of the 
       normalisation procedure. 
       The dotted vertical lines indicate the rest-frame velocity of the host galaxy.}
              \label{fig:broad}
    \end{figure}

The spectra of SN~2013gh display the deep multiple-feature Na~I~D and Ca~II~H\&K lines shown in Figure~\ref{fig:naid}. 
The redshifted wavelength of Ca~II~K nearly coincides with the rest-frame wavelength of Ca~II~H, 
causing the host-galaxy Ca~II~K and Milky Way Ca~II~H lines to blend. At the host-galaxy rest-frame wavelength 
we detect the DIBs listed in Table~\ref{tab:ew_13gh}, 
of which the DIBs $\lambda\lambda$5797, 6379, and 6614 are most clearly defined and plotted in Figure~\ref{fig:dib}. 
Unfortunately, the DIB~$\lambda$5780, which was discussed as a reddening proxy by \citet{2013ApJ...779...38P}, 
falls into a gap not covered by the UVES spectra.

In Table~\ref{tab:ew_13gh} the measured EW values of all identifiable ISM species of SN~2013gh are presented.
We use the relations of \citet{2012MNRAS.426.1465P} and \citet{2015MNRAS.447..545B} to estimate 
$E(B-V)$ from Na~I~D and the DIB~$\lambda 5797$, respectively. 
It can be visually seen that the line ratios Na~I~D2/D1 are $<2$, implying that the \citet{2012MNRAS.426.1465P}
relation at best gives a lower limit for $E(B-V)$.
Nevertheless, the $E(B-V)$ limit alone places SN~2013gh into the group of SNe with unusually 
high Na~I column densities as defined by \citet{2013ApJ...779...38P}.
The value of $E(B-V)$ can furthermore be estimated from the EW of 
DIBs using empirical relations \citep[e.g.,][]{2015MNRAS.447..545B}.
The DIB~$\lambda5797$, for example, is roughly consistent with the measured $E(B-V)$ from photometry.
Interestingly, most of the DIBs are comparable within uncertainties to those detected in 
SN~2001el \citep{2005A&amp;A...429..559S}, a SN~Ia that matches SN~2013gh in
photometric reddening.

It can be seen in Figures~\ref{fig:naid}~and~\ref{fig:dib}, that the typical ISM features 
appear to be redshifted with respect to the recession velocity of NGC~7183. 
Furthermore, the coordinates of the SN are consistent with a location 
in the redshifted arm in Very Large Array (VLA) H~I data\footnote{D. Vergani, 2015, private communication.}. 
The deepest absorption features in both Na~I~D and Ca~II~H 
span velocities of $v\approx10$--$30$km~s$^{-1}$. Another blended set 
of features appears redshifted from these at $v\approx40$km~s$^{-1}$.
Finally, a small, seemingly unblended feature appears at $v\approx73$km~s$^{-1}$ in Na~I~D, but it is absent from Ca~II~H\&K.

On the blueshifted side of the Na~I~D profile, there appears to be a shallow diffuse feature or wing of a Voigt profile.
In Figure~\ref{fig:broad}, we show the non-normalised spectra at the three epochs around the Na~I~D
doublet. The broad feature can be visually identified at all three epochs 
and is thus not an artefact of the normalisation procedure.
We find that the feature is not symmetric about the deepest features of the Na~I~D profile, 
indicating that it may be a broad diffuse component of the ISM.
In Figure~\ref{fig:broad}, broad Gaussians with $\sigma_{\rm vel}\approx50$km~s$^{-1}$ are shown,
which are centred on the rest-frame velocity of NGC~7183.

Interestingly, an H~I spectrum extracted at the position of SN~2013gh exhibits emission which is centred
on and spans approximately the same range as the broad feature. 
It is therefore possible that the broad feature corresponds to diffuse Na~I gas tracing H~I.
If this is the case, then the broad feature corresponds to the warm ISM, unlike 
the deep Na~I~D absorption features which typically trace the cold ISM.
Assuming that the broad feature traces H~I gas, this could  
suggest that SN~2013gh is located on the far side of NGC~7183.


\subsection{Varying Na~I~D Feature}
\label{ssec:change}

   \begin{figure*}
   \centering
       \resizebox{\hsize}{!}{\includegraphics{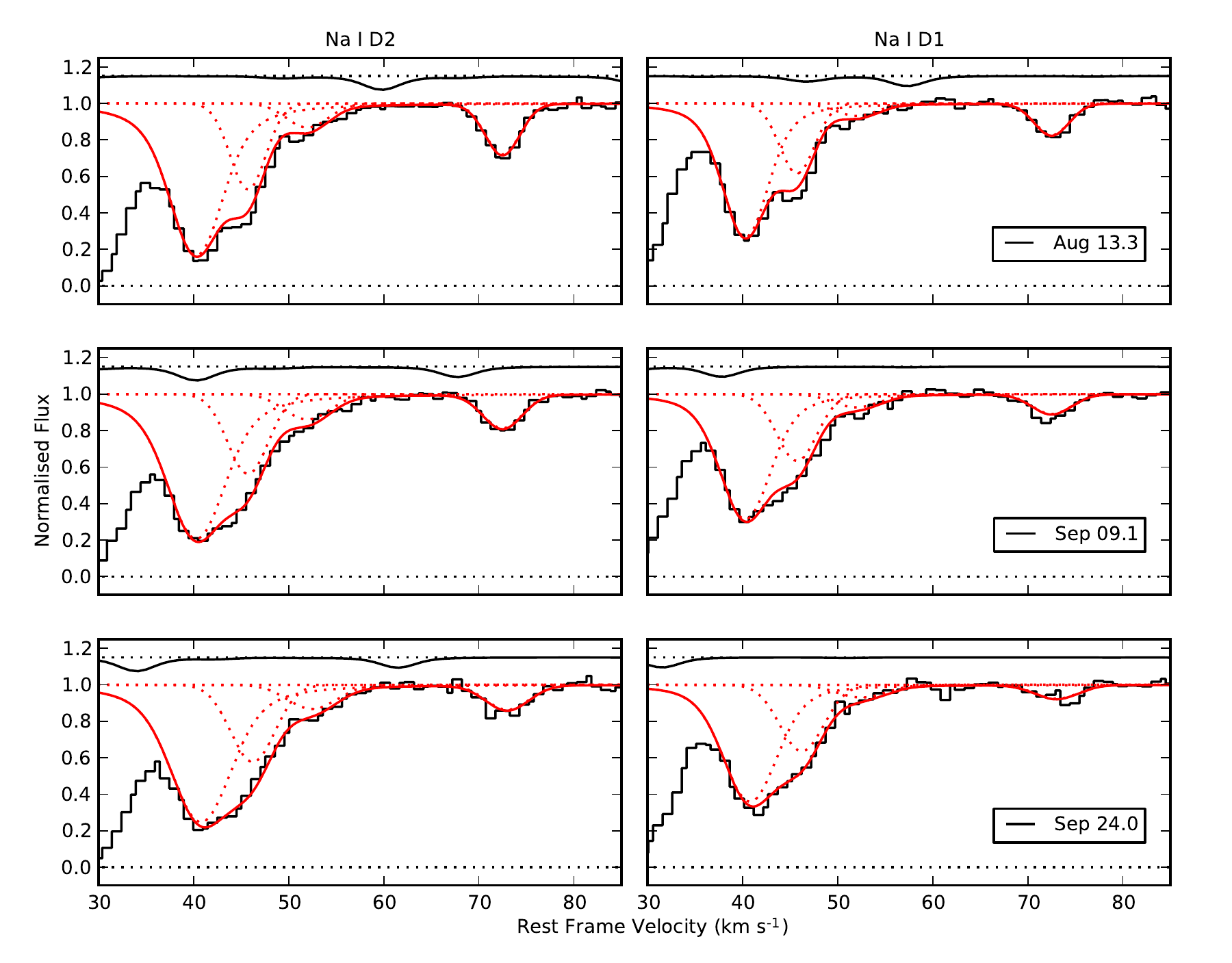}}
       \caption{The redshifted features of the Na~I~D profile of SN~2013gh along with the Voigt profiles obtained with VPFIT.
       The model telluric spectra of the respective epochs are plotted for reference above the continua.
       It can be seen that the feature at $v\approx73$~km~s$^{-1}$ probably changes in depth, 
       whereas the other features only appear to change because of differences in the resolution of the spectra.
       }
              \label{fig:zoom}
    \end{figure*}

\begin{table*}
  \centering
  \begin{tabular}{l r@{\,}l r@{\,}l r@{\,}l r@{\,}l r@{\,}l r@{\,}l r@{\,}l r@{\,}l}
    \hline\hline
    & \multicolumn{8}{c}{\small SN~2013gh components} \\
    \hline
    \small{$v$ (km~s$^{-1}$)} & {\small 73.0} & {\small $\pm$ 0.1} & {\small 52.4} & {\small $\pm$ 0.6} & {\small 46.2} & {\small $\pm$ 0.3} & {\small 40.8} & {\small $\pm$ 0.1} \\
    \small{$b$ (km~s$^{-1}$)} & {\small 0.76} & {\small $\pm$ 0.14} & {\small 1.88} & {\small $\pm$ 0.84} & {\small 0.74} & {\small $\pm$ 0.06} & {\small 0.39} & {\small $\pm$ 0.07} \\
    \hline
     \small{Epoch}  & \multicolumn{8}{c}{\small ${\rm log}_{10}\{N_{\rm Na~I}\,($cm$^{-2})\}$} \\
    \hline
    	{\small Aug.~13.3} & {\small 11.36} & {\small $\pm$ 0.05} & {\small 10.95} & {\small $\pm$ 0.07} & {\small 11.95} & {\small $\pm$ 0.06} & {\small 14.73} & {\small $\pm$ 0.03} \\
	{\small Sep.~9.1} & {\small 11.18} & {\small $\pm$ 0.04} & {\small 11.02} & {\small $\pm$ 0.05} & {\small 12.05} & {\small $\pm$ 0.06} & {\small 14.75} & {\small $\pm$ 0.02} \\
	{\small Sep.~24.0} & {\small 11.05} & {\small $\pm$ 0.04} & {\small 11.03} & {\small $\pm$ 0.07} & {\small 12.17} & {\small $\pm$ 0.09} & {\small 14.73} & {\small $\pm$ 0.03} \\
    \hline\hline
    \multicolumn{1}{l}{}\\
  \end{tabular}
  \caption{Voigt profile parameters of the redshifted features of Na~I~D in SN~2013gh determined with VPFIT. 
  Profiles are displayed in Figure~\ref{fig:zoom}.
  The velocity ($v$) is with respect to the rest frame of NGC~7183. 
  The Doppler broadening ($b$) has been determined from the first epoch and fixed in the following two. 
  \label{tab:voigt_13gh}}
\end{table*}

To search for line variations in SN~2013gh, we have measured the EWs of all features and compared the individual epochs.
While DIBs and Ca~II~H\&K appear not to change significantly, 
there is a slight decrease in the EW of Na~I~D.
Guided by visual inspection, there appear to be differences between the individual epochs in the redshifted part of the Na~I~D profile.
In Figure~\ref{fig:zoom}, the profiles of the redshifted Na~I~D features of each epoch and the fitted Voigt profiles are shown.
The positions of telluric lines can be inferred from the model telluric spectra plotted above the respective SN spectra.

The higher resolution of the first epoch is evident in the region $v=40$--$50$km~s$^{-1}$, 
where the troughs between individual components can be visually distinguished.
Estimates of the resolution ($R = \lambda/\Delta \lambda)$ for each spectrum can be found in Table~\ref{tab:spec}. 
We find that the feature at $v=35$--$60$~km~s$^{-1}$ can be fit by three Voigt profiles.
The EW in this velocity range does not change significantly between the individual epochs
and the column densities of the individual components inferred from the Voigt profile fits
do not show any significant variations either.
The Voigt profile parameters of the best fit are summarised in Table~\ref{tab:voigt_13gh}.
Since neither the column densities nor the total EW of the components at $v=35$--$60$km~s$^{-1}$ appear to vary,
the visual differences between the individual epochs must be caused by differences in the resolution of the spectra.

The most redshifted feature at $v\approx73$km~s$^{-1}$, however, does appear to vary between the epochs.
It is important to ascertain that the variations are not an effect of changing telluric lines.  
First, the feature 
in both Na~I~D1 and D2 display the same trend, implying that telluric lines would need to overlap with both features
to explain the change. Second, the depth of the telluric lines would have to change with the same trend as 
observed in the spectra. In Figure~\ref{fig:zoom}, the model telluric spectra of the individual epochs are plotted to show the 
approximate locations of telluric lines with respect to the feature. 
No telluric lines appear to overlap with the feature in any of the epochs, 
nor is there an obvious trend in the depth of the telluric lines.
Furthermore, we ascertained that the depth of the telluric lines in the standard-star spectra obtained on the 
same epochs do not display any trend either.
Thus, poorly subtracted telluric lines are unlikely to account for the variations observed in the feature at $v\approx73$km~s$^{-1}$.

We further checked that the changes are not an unusual systematic error, 
in which case other absorption features might display a similar trend. 
The other components of the Na~I~D doublet of the host 
galaxy are much deeper than the feature to which we would like to compare them. However, the Milky Way Na~I~D
doublet consists of three components of comparable depth to the changing feature. The EWs of 
these three Milky Way features neither show any significant variations between the epochs nor any trend.
Lastly, individual exposures of the same epochs (see Figure~\ref{fig:all_exp}) agree very well with each other,
indicating that there are no systematic changes within a night.
From the above we conclude that the reduced and normalised spectra can be trusted.

We have measured the EWs of the feature at $v\approx73$km~s$^{-1}$ in each exposure.
The measurements are presented in Appendix~\ref{ap:bootstrap}, and we also convince ourselves that the 
decrease in measured EW is significant.
This was done using a bootstrap analysis that indicates that the change in EW is inconsistent with 
random noise of a feature which does not vary with time. 

The Voigt profile fits further confirm the assertion that the $v\approx73$~km~s$^{-1}$ feature is changing.
Interestingly, the Doppler broadening of the feature was found to be 
much more narrow than the observed profile with $b=0.8\pm0.1$~km~s$^{-1}$.
Thus, the feature is unresolved, deeper than it appears, and in fact not optically thin. 
Visually, the varying feature appears to only decrease in depth and not in width, which is consistent with an unresolved line profile.
In Table~\ref{tab:voigt_13gh} the Na~I column densities inferred from the Voigt profile fits are presented.
The Doppler broadening ($b$) of the respective features were determined from the first epoch and fixed for the following epochs.
The fits indicate that the column density of the $v\approx73$km~s$^{-1}$ feature is decreasing with time, while the other features are not changing.
Although the changing feature could also be explained by a decrease in intrinsic line width with unchanging column density,
there is no obvious physical scenario how this could be the case.
Thus, the best explanation for the changes seen in the data is a decrease in column density.

To summarise, we have identified an Na~I~D component at $v\approx73$~km~s$^{-1}$ which appears to be varying with time.
Telluric lines cannot account for the variations, since there are no telluric lines that overlap with the changing feature.
Furthermore, there does not appear to be any reason to believe that the variations might be caused by some systematic effect.
Individual exposures at the same epochs seem to be very consistent with one another and no other features show the same 
trend over time.
The decreasing EW can be measured in all individual exposures of the three epochs, and a simple bootstrap test has shown that 
it is very unlikely to obtain this result from random noise.
The Voigt profile fits further confirm that the column density of Na~I is decreasing. 
Also, the fits demonstrate that the line is very narrow and the profile is in fact not resolved.
We thus conclude that the variations seen in the feature are real 
and most likely produced by a decreasing column density of Na~I atoms along the line of sight.

\subsection{iPTF~13dge}

\begin{figure*}
   \centering
        \resizebox{\hsize}{!}{\includegraphics{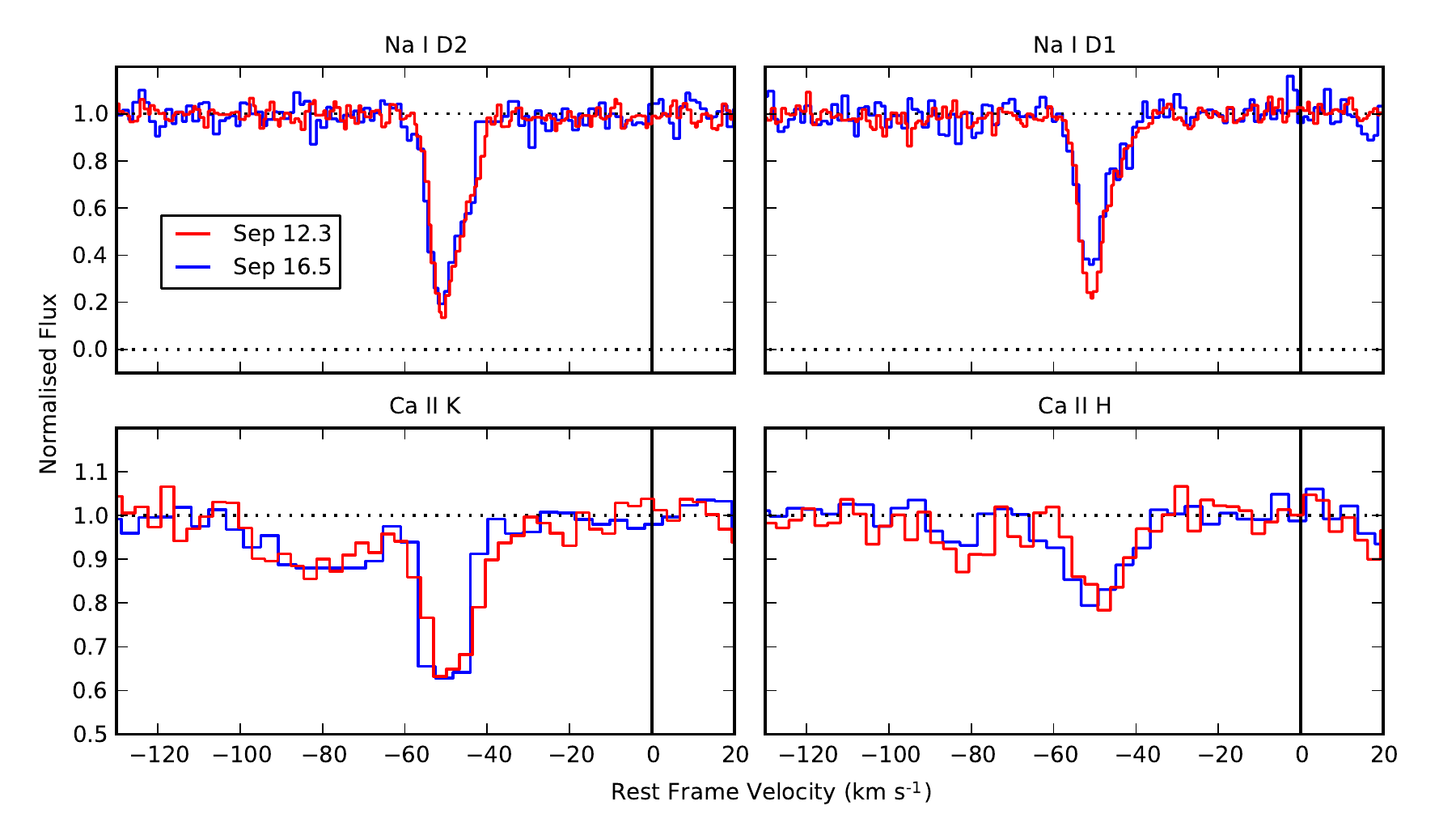}}
       \caption{Na~I~D  and Ca~II~H\&K of iPTF~13dge obtained with UVES (Sep. 12.3) and HIRES (Sep. 16.5). 
       The Ca~II~K\&H lines have been binned for clarity.
       Cosmic rays in Na~I~D2 and Ca~II~K in the HIRES spectra have been masked.}
              \label{fig:dge}
    \end{figure*}

The two high-resolution spectra obtained of iPTF~13dge were taken only four nights apart.
However, the early epochs at which the spectra were taken make the interval sensitive to 
photoionisation 
as will be shown in Section~\ref{ssec:absorbers}. 
The HIRES spectrum is contaminated by many cosmic rays, 
affecting both the Na~I~D and Ca~II~H\&K lines.
The Na~I~D doublets as well as Ca~II~H\&K of both spectra are shown in Figure~\ref{fig:dge}, though cosmic rays were masked in the HIRES spectrum.
Interestingly, the Ca~II lines appear to have an additional blueshifted feature not 
visible in the Na~I~D doublet. No DIBs could be identified in either spectra.
The EW values of the detected features are given in Table~\ref{tab:dge} along with $E(B-V)$ estimates 
using the \citet{2012MNRAS.426.1465P} relation.
The Na~I~D2/D1 line ratio clearly shows that the lines are not optically thin. 
This in turn implies that the $E(B-V)$ estimates are merely lower limits.
Although the value is in good agreement with the $E(B-V)$ presented in Table~\ref{tab:lcs}, 
one must keep in mind that the photometric value has been computed with a fixed $R_{V}=3.1$.

Comparing the absorption lines of both epochs has not revealed any significant variations.
The Ca~II~H\&K EWs are consistent with one another within the uncertainties.
We inspected the Na~I~D doublet in more detail, 
because there seems to be a slight difference between the epochs in Na~I~D2. 
While the EW of Na~I~D2 does not change between the epochs, 
there appears to be a $2\sigma$ decrease in EW in Na~I~D1.
Since this could be consistent with a decreasing feature in the nonlinear part of the curve of growth, 
we fit Voigt profiles to the features to determine the Na~I column densities.
The Na~I~D profile is consistent with two components, the parameters of which are presented in 
Table~\ref{tab:voigt_13dge}.
Unfortunately, the column densities obtained from the Voigt profiles do not change significantly.
Thus, no definite change could be detected in any absorption feature of iPTF~13dge.

    \begin{table}
       	\centering
  	\begin{tabular}{l r@{\,}l c}
    	\hline\hline
    	{\small Feature}  &  \multicolumn{2}{c}{\small EW} & {\small Estimated $E(B-V)$}\\
	& \multicolumn{2}{c}{\small (m\AA)} & {\small (mag)}\\
    	\hline
	{\small Na~I~D1} & {\small 150} & {\small $\pm$ 6} & {\small >0.04$^{+0.02}_{-0.01}$}\\       
	{\small Na~I~D2} & {\small 175} & {\small $\pm$ 6} & {\small >0.03$^{+0.01}_{-0.01}$}\\       
	{\small Ca~II~K} & {\small 134} & {\small $\pm$ 10} & {\small - }\\
	{\small Ca~II~H} & {\small 65} & {\small $\pm$ 11} & {\small - }\\
	\hline\hline
	\multicolumn{1}{l}{}\\
  	\end{tabular}
  	\caption{The equivalent widths of Na~I~D and Ca~II~H\&K of iPTF~13dge. 
	The $E(B-V)$ is computed using the \citet{2012MNRAS.426.1465P} empirical relation. 
	\label{tab:dge}}
\end{table}

\begin{table}
  \centering
  \begin{tabular}{l r@{\,}l r@{\,}l r@{\,}l r@{\,}l}
    \hline\hline
    & \multicolumn{4}{c}{\small iPTF~13dge components} \\
    \hline
    \small{$v$ (km s$^{-1}$)} & {\small -50.8} & {\small $\pm$ 0.2} & {\small -45.3} & {\small $\pm$ 0.5} \\
    \small{$b$ (km s$^{-1}$)} & {\small 0.95} & {\small $\pm$ 0.11} & {\small 3.5} & {\small $\pm$ 0.8} \\
    \hline
     \small{Epoch}  & \multicolumn{4}{c}{\small $log_{10}\{N_{Na~I}\,($cm$^{-2})\}$} \\
    \hline
    	{\small Sep.~12.3} & {\small 13.82} & {\small $\pm$ 0.45} & {\small 11.64} & {\small $\pm$ 0.09} \\
	{\small Sep.~16.5} & {\small 13.64} & {\small $\pm$ 0.15} & {\small 11.63} & {\small $\pm$ 0.04} \\
    \hline\hline
    \multicolumn{1}{l}{}\\
  \end{tabular}
  \caption{Voigt profile parameters of Na~I~D in iPTF~13dge determined with VPFIT.
  The velocity ($v$) is with respect to the rest frame of NGC~1762.
  The Doppler broadening ($b$) has been determined from the first epoch and fixed in the second epoch. 
  \label{tab:voigt_13dge}}
\end{table}


\section{Interpretation of Varying Absorption Lines}
\label{sec:ion}

In Section~\ref{ssec:change}, we have seen that the small redshifted feature of the 
Na~I~D doublet of SN~2013gh is varying over time.
There are several possibilities by which this could occur.
The first two we consider are geometric effects caused by either rapidly moving or patchy clouds along the line of sight.
Another possibility is that 
photoionisation produced by the UV flux of the SN decreases the column density of an absorber.

\subsection{Geometric Effects}

A variety of geometric effects could give rise to varying column densities of absorbers.
Unfortunately, the varying Na~I~D feature is not detected in the profiles of any other ISM species for comparison.
The absence of the feature in Ca~II~H\&K suggests that there is very little 
Ca~II in the cloud or it is locked up in dust grains.
Thus, it is not possible to use Ca~II~H\&K to test for a geometric origin of the variations seen in Na~I~D.

One geometric effect we considered is that the small feature corresponds 
to a gas cloud passing through the line of sight to the SN. 
To observe variations on the timescale of weeks, a gas cloud must have column density variations on small scales 
and be moving at a high velocity.
The size of the column density variations and the velocity of the cloud are inversely proportional.
For instance, one could suppose that the small varying feature has a transverse velocity 
of the same order of magnitude ($v_{\rm trans}\approx100$~km~s$^{-1}$) as the line-of-sight velocity with respect to the galaxy rest frame.
To then see variations on the order of tens of days, the gas cloud must have a diameter of scale $R_{\rm trans}\approx1$ AU.

A second possibility is that an unchanging gas cloud is close enough to the SN 
for there to be a projection effect on the expanding photosphere \citep{2010A&A...514A..78P}.
The scenario is possible if, for example, a small gas cloud initially overlaps with the entire photosphere, 
but later is too small to veil the expanded photosphere.
The method is restricted to detecting column density variations in gas clouds 
that are smaller in scale than the size of the photosphere.
Assuming a spherical gas cloud with homogeneous volume density  and centred on the photosphere, 
the variations in Na~I~D of SN~2013gh are consistent with a cloud 
having a radius $R_{\rm proj}\approx220$ AU.
This would imply that the true geometry of the gas cloud must vary on scales smaller than $R_{\rm proj}$.

The two geometric effects considered above could be the origin of the variations detected in Na~I~D.
However, this could only be confirmed
if the corresponding absorption feature were detected in another absorption profile
and displayed exactly the same trend in column density over time.
As in the case of SN~2006X, a geometric origin of the Na~I~D 
variations in SN~2013gh cannot be ruled out \citep{2007Sci...317..924P}.

\subsection{Photoionisation}
	\label{ssec:ion}

Another possible source of absorption-line variations is by photoionisation of the gas by UV radiation from the SN.
Knowing the UV flux of a SN and the ionisation cross-section of an absorber, 
the ionisation rate at a given distance from the SN and at a given phase 
can be modelled \citep{2009ApJ...699L..64B}.

The biggest uncertainty in the model is the UV flux of the SN itself.
In particular, only a few SNe have had UV spectra taken at early phases.
Furthermore, it is known that the largest differences between 
individual SNe occur in the UV part of the spectrum \citep[see, e.g.,][]{2010ApJ...721.1608B}.
This becomes apparent when comparing the commonly used templates 
of \citet{2007ApJ...663.1187H} (hereafter H07) and SN~2011fe \citep[e.g.,][]{2015MNRAS.453.3300A}.
Below $\sim2400$~\AA, the ionisation energy of Na~I \citep[via NIST;][]{NIST_ASD}, the peak 
flux of the H07 template is a factor of $\sim4$ greater than that of SN~2011fe.

Motivated by the excellent UV coverage and low extinction of SN~2011fe, 
we used the spectral energy distribution~(SED) 
described by \citet{2015MNRAS.453.3300A} to compute the photoionisation rates.
However, the H07 template extends to both shorter wavelengths and earlier phases than the SN~2011fe template.
We therefore extrapolated the SN~2011fe SED by that of H07, 
which was scaled to match the flux at the ``edges'' of the SN~2011fe template at $1750$~\AA .
Furthermore, the early phases of the H07 template were scaled by 
a factor determined from the earliest SN~2011fe spectrum.

   \begin{figure}
   \centering
       \resizebox{\hsize}{!}{\includegraphics{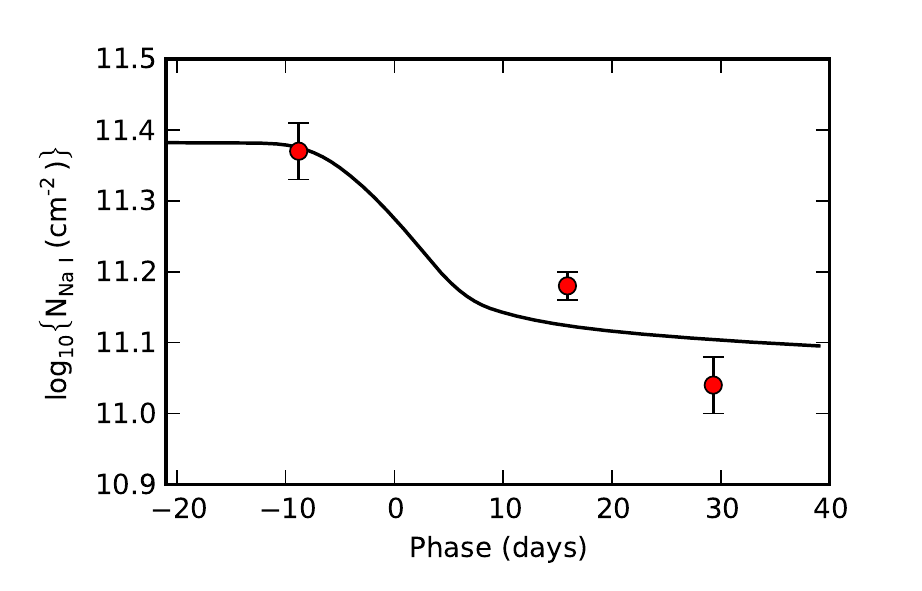}}
       \caption{The best-fit photoionisation curve for the varying Na~I~D feature in SN~2013gh.
       }
              \label{fig:photoion}
    \end{figure}

Using the extrapolated SED and wavelength-dependent ionisation cross-sections
\citep{1996ApJ...465..487V}\footnote{Obtained via \url{phidrates.space.swri.edu}}, 
we can model the photoionisation 
of an absorber. Assuming that the gas is optically thin and situated in a thin shell,
the ionisation rate is characteristic for the radius of the shell.
Therefore, we can fit the model to the column-density data obtained from the changing Na~I~D 
feature of SN~2013gh to determine the initial column density of the gas and the distance from the SN.
In Figure~\ref{fig:photoion}, the best-fit photoionisation curve is plotted along with the Na~I 
column-density measurements of the varying feature.
For the best fit, we find that the initial column density must have been $\log_{10}\{N_{\rm Na~I}\,($cm$^{-2})\} = 11.38\pm0.02$
at a distance $\log_{10}\{R_{\rm ion}\,($cm$)\} = 18.9\pm0.2$, 
where the uncertainties incorporate a range of total ionising flux of $\pm1$~mag, 
to crudely estimate the effects of different UV brightness.

We have independently verified the photoionisation result with an ionisation and excitation code used to study 
absorption-line variability in gamma-ray burst afterglow environments from \citet{2013A&amp;A...549A..22V}.
The resulting radius $\log_{10}\{R_{\rm ion}\,($cm$)\} = 18.89\pm0.04$ and initial column density 
$\log_{10}\{N_{\rm Na~I}\,($cm$^{-2})\} = 11.38\pm0.04$ are in excellent agreement with the above results.
The smaller uncertainty in the distance does not take into account the possible range of flux of the SED.

Since we consider photoionisation as a possible cause for the observed line variation, 
we would like to argue why recombination is not relevant on the timescale of the observations.
Following a similar a argument as by \citet{2009ApJ...702.1157S}, the recombination timescale is
\begin{equation}
\tau_{\rm rec}=\frac{1}{\alpha n_{e}},
\end{equation}
where $\alpha$ is a rate coefficient for a given temperature and $n_{e}$ is the volume density of electrons.
Since the properties of the varying cloud are not known, we will determine a conservative value of $n_{e}$ 
for a recombination timescale which would affect the above results. 
Assuming a typical ISM cloud temperature of $\sim10^2$~K \citep{2012ApJ...756..157G},
the rate coefficient for Na~I is $\alpha\approx10^{-11}$~cm$^3$ s$^{-1}$ \citep{2006ApJS..167..334B}.
Note that this assumption is conservative; the recombination time will increase with greater temperature,
if the gas is heated by the SN or progenitor. 
Recombination would need to be taken into account if the recombination timescale 
is $\sim10$~days or $\tau_{\rm rec}\approx10^6$~s.
This would imply an electron density $n_e \approx 10^5$~cm$^{-3}$.
\citet{2008AstL...34..389C} has used this $n_e$ value for electrons streaming from an SD progenitor 
at $\sim 10^{17}$ cm from the SN. 
At greater distances, however, the electron density should drop off,
at $10^{19}$ cm being much lower than the limit computed.
Thus, recombination should be negligible in the case of the varying Na~I~D feature, or else
the ionisation rate is underestimated by the simple photoionisation model.

\subsection{Absorbers in the CS Environment}
\label{ssec:absorbers}

   \begin{figure}
   \centering
       \resizebox{\hsize}{!}{\includegraphics{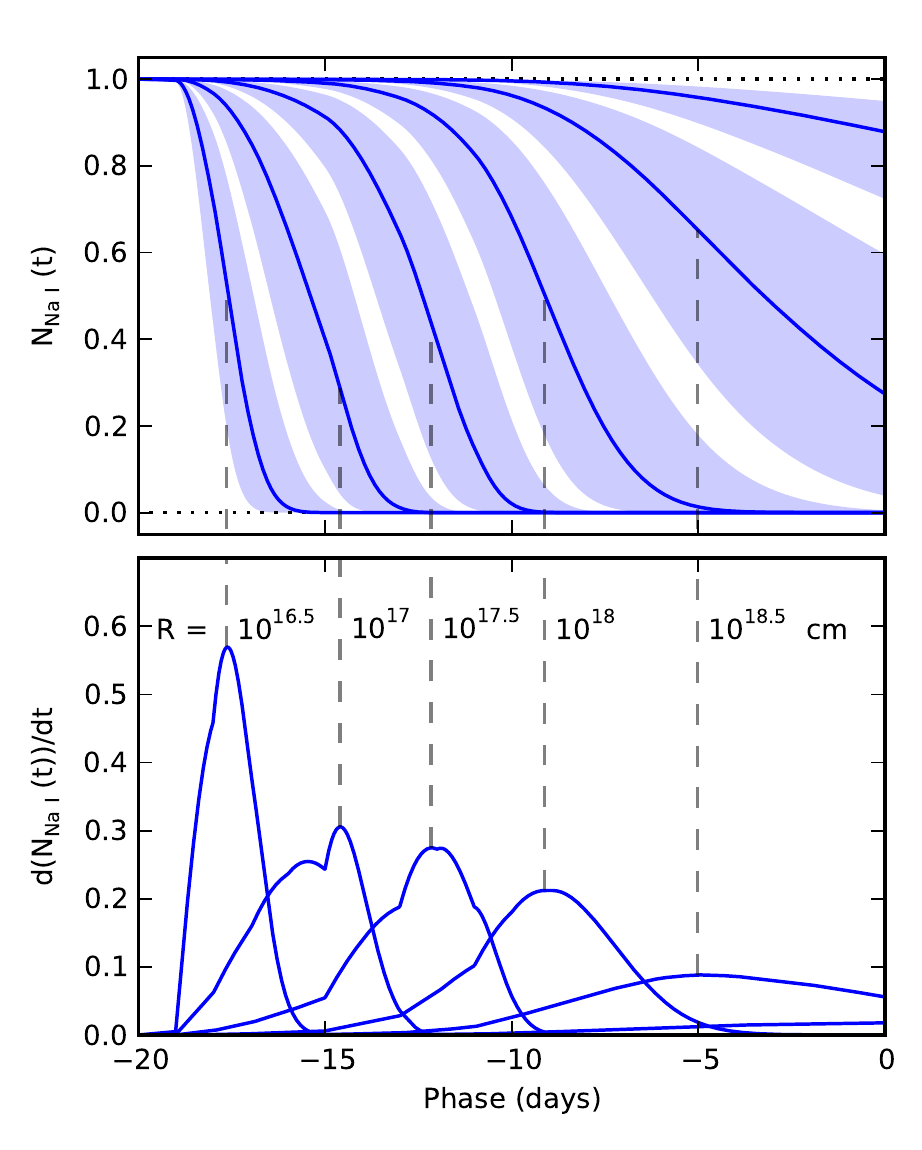}}
       \caption{The Na~I photoionisation fraction and rate as a function of phase at distances 
       between $10^{16.5}$ and $10^{18.5}$ cm from a SN.
       The dashed vertical lines associate the curves of the two plots with one another 
       and are labeled by their respective radii at which the gas is situated from the explosion.
       The shaded bands indicate the $\pm1$~mag change in ionising UV flux from the SN.
       The Na~I gas is assumed to be optically thin and is completely ionised at all radii considered in this plot.
              \label{fig:ion_limits}}
    \end{figure}

   \begin{figure}
   \centering
       \resizebox{\hsize}{!}{\includegraphics{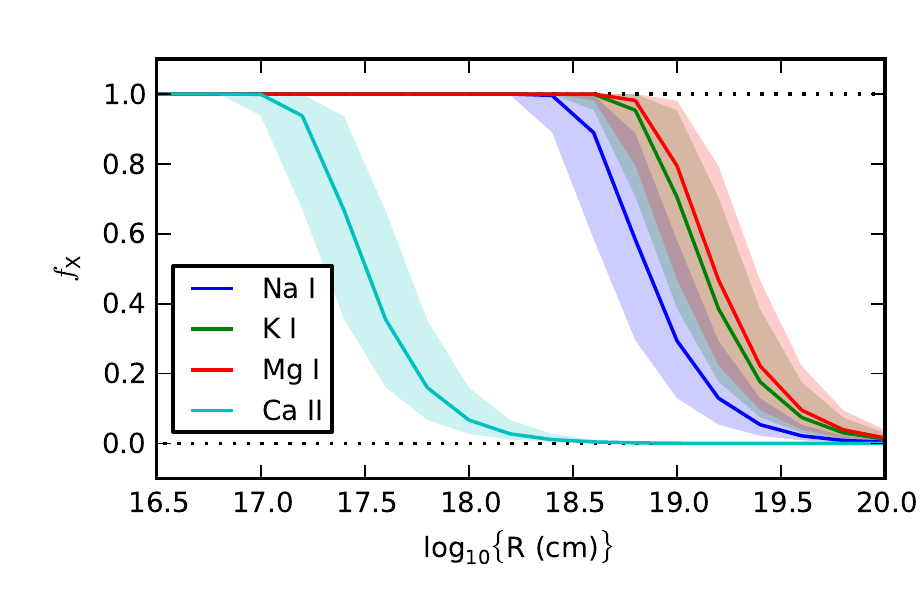}}
       \caption{The total fraction of photoionisation of different ISM species as a function of distance from the SN.
       The bands indicate a $\pm 1$ mag change in ionising UV flux.
       }
              \label{fig:ion_species2}
    \end{figure}

   \begin{figure}
   \centering
       \resizebox{\hsize}{!}{\includegraphics{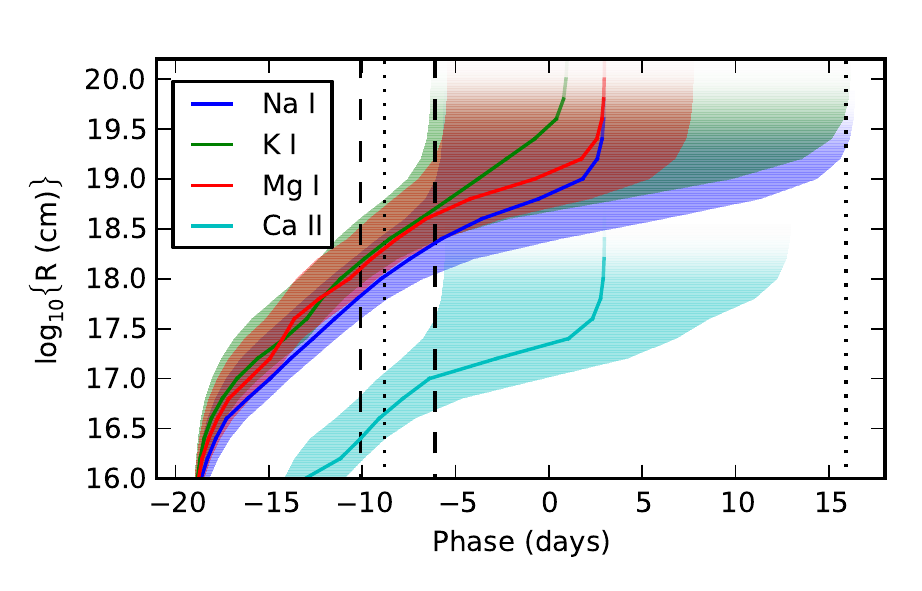}}
       \caption{The phases at which photoionisation occurs for different ISM species as a function of radius.
       The lines correspond to the peak ionisation rate, and the shaded region indicates the phases 
       between which $80\%$ of photoionisation occurs. The fading corresponds to the limit 
       where $<5\%$ of the gas is ionised in total.
       The $\pm 1$ mag range of the ionising flux is \textit{not} considered in this plot. 
       The phases of the spectra taken of SN~2013gh (dotted lines) and iPTF~13dge (dashed lines) indicate 
       which radii are probed for photoionisation.
       }
              \label{fig:ion_species1}
    \end{figure}

    \begin{table}
       	\centering
  	\begin{tabular}{c@{\,}c c c}
    	\hline\hline
	&& {\small Ionisation} & {\small Ionisation}\\
    	\multicolumn{2}{c}{\small Species}  &  {\small Energy}  & {\small Cross-section$^{\dagger}$}\\
	&& {\small [eV]} & {\small [cm$^2$]}\\
    	\hline
	{\small Na}&{\small ~I}& {\small 5.14} & {\small $7.7\times10^{-20}$}\\
	{\small Mg}&{\small ~I} & {\small 7.65} & {\small $3.0\times10^{-18}$}\\
	{\small K}&{\small ~I} & {\small 4.34} & {\small $8.9\times10^{-21}$}\\
	{\small Ca}&{\small ~II} & {\small 11.9} & {\small $5.4\times10^{-19}$}\\
	\hline\hline
	\multicolumn{3}{l}{$^{\dagger}$At ionisation energy.}\\
  	\end{tabular}
  	\caption{Ionisation energies and cross-sections of some typical ISM species. 
	\label{tab:ions}}
\end{table}

The photoionisation rate is a function of the shell radius of an absorber and of the time after the explosion.
In Figure~\ref{fig:ion_limits}, the time-dependent photoionisation of Na~I gas of shells at 
radii ranging from $10^{16.5}$ to $10^{18.5}$ cm are considered.
The upper plot shows the fractional decrease of Na~I with time with respect to the initial column density.
In the lower plot the rate of photoionisation of the respective gas shells is shown.
To probe for photoionisation at a given radius from the SN it is thus necessary to obtain spectra spanning 
the time during which there is a significant rate of photoionisation.
For instance, one must obtain spectra before $\sim8$ days prior 
to maximum light to probe radii of $< 10^{18}$ cm using Na~I absorption.
The exact phase at which the maximum rate of photoionisation occurs depends on the UV luminosity of the SN. 
The brighter a SN is in UV radiation, the earlier ionisation will occur. 
In Figure~\ref{fig:ion_limits}, the shaded bands show the range of ionisation curves 
if the brightness is varied by $\pm1$~mag.
The model does not take into account 
recombination (see Section~\ref{ssec:ion}),
preionisation by the progenitor system 
or shock-breakout UV flash \citep{1996MNRAS.283.1355C}, which could ionise the 
innermost radii from the SNe.

From the absence of absorption-line variations, the photoionisation model can also be used to 
exclude radii at which the absorber is located from the SN.
Even though the high-resolution spectra of iPTF~13dge were taken only 4 days apart, 
it can be seen in Figure~\ref{fig:ion_limits} that significant ionisation can occur during these phases.
Between $-9$ and $-5$ days before maximum brightness, the photoionisation model is sensitive to 
ionisation occurring at distances of $6\times10^{17}$--$5\times10^{18}$ cm from the SN, 
whereby the possible range in UV brightness is not taken into account.
It thus indicates that the visible Na~I~D features are not located at these radii.

So far the search for varying absorption lines has mainly focused on the Na~I~D doublet. 
This is motivated by fact that it is relatively easy to obtain high-S/N spectra around the wavelength of Na~I~D.
Being a doublet furthermore helps to ascertain whether variations in Na~I~D are real, 
when they are detected in both profiles.
The recent example of varying K~I lines in SN~2014J \citep{2015ApJ...801..136G} has shown that it is not 
necessary to solely focus on Na~I~D.
We therefore investigated some other typical ISM species.

In addition to Na~I, we have investigated the photoionisation of 
the absorbers in Table~\ref{tab:ions} more closely.
While the ionising fluxes of Na~I, K~I, and Mg~I are comparable, that of Ca~II is roughly two orders of magnitude lower. 
The SED used does not cover wavelengths $< 1000$~\AA, implying that some ionising flux is not taken into consideration.
The SED cutoff is inadvertently close to the ionisation energy of H~I at $911$~\AA,
beyond which the H~I gas likely shields all other particles from ionising flux.
Judging by the rapid drop in flux at higher energies, the flux beyond the cutoff would only contribute a few percent to 
the total ionising flux of Na~I, K~I, and Mg~I.
Unfortunately, the ionisation energy of Ca~II at $1044$~\AA\ is close to the cutoff, and thus the total 
ionising flux is underestimated. 
Assuming a roughly constant continuum in the range $911$--$1044$~\AA, the ionising flux of Ca~II could be a factor $\sim3$
greater than the flux computed from the SED.
Hence, the details of the ionising effects on Ca~II must be taken with a grain of salt, but are presented below since they show that 
photoionisation of Ca~II can occur around SNe~Ia.

In Figure~\ref{fig:ion_species2}, the final ionisation fraction as a function of radius is shown for the different absorbers.
While the total ionisation of Na~I, Mg~I, and K~I drop off at similar radii around $\sim10^{19}$~cm, 
Ca~II is not ionised beyond a radius of $\sim10^{18}$~cm owing to the high ionisation energy.
Figure~\ref{fig:ion_species1} shows which radii are probed for photoionisation by taking spectra at different phases.
From Figure~\ref{fig:ion_species1}, one can infer to which radii the observations presented in this paper were sensitive.
The large time interval between the first and second epoch of SN~2013gh spectra implies that a large range of radii
$10^{17.5}$--$10^{19.5}$~cm has been probed with Na~I.
The short interval between the two iPTF~13dge epochs, on the other hand, implies that the smaller range of radii 
$10^{18}$--$10^{18.5}$~cm has been probed.
Although similar limits could be set using Ca~II, the uncertainty of the UV flux beyond 
the reaches of the SED would probably make them unreliable.

\section{Summary and Conclusions}
\label{sec:discussion}

Our high-resolution spectra of the SN~2013gh and iPTF~13dge can be 
added to the slowly growing sample of narrow absorption line temporal series of SNe~Ia.

SN~2013gh is a reddened SN with negligible Galactic extinction ($E(B-V)_\mathrm{MW}=0.025$~mag) and 
with a line of sight close to a dust lane in the host galaxy NGC~7183.
Comparison of the Na~I~D and H~I profiles indicates that SN~2013gh likely occurred on the far side
of its host, indicating that the reddening is mainly due to dust in the host galaxy.

The time series of Na~I~D absorption has revealed a small redshifted feature with decreasing column density. 
Although the change is consistent with photoionisation occurring at $\sim10^{19}$ cm from the explosion,
geometric effects cannot be excluded since the feature is not discernible in any other absorption-line profile.
The time-varying component is unusual in that it is the most redshifted feature in the Na~I~D profile.
If the feature is associated with outflowing gas from the progenitor,
the SN could be located in a part of the host galaxy that is receding faster than the main absorption lines.
However, the gas could also be unrelated to the SN and moving toward it.
Unfortunately, the crowded line of sight close to the core of NGC~7183 and 
the proximity to a dust lane make a more definite determination of the origin of the varying Na~I~D 
feature in SN~2013gh difficult.

The temporal coverage of the spectra we have analysed should be sensitive to photoionisation of Na~I gas occurring 
$10^{17.5}$--$10^{19.5}$ cm from the SN, thus probing radii that can be considered part of the CS environment. 
The nondetections of further time-varying absorption features indicate that the high Na~I column density in the 
SN~2013gh spectra does not originate from gas in these regions.
Within $10^{17.5}$ cm, all Na~I gas would have been ionised before the first spectrum was taken.
Thus, it appears that the bulk of the gas must be situated in the ISM beyond $10^{19.5}$ cm from the SN.
Compared to the reddening, the column density of Na~I gas also appears to be unusually high
with respect to the \citet{2012MNRAS.426.1465P} relation.

It should also be possible to exclude the presence of Ca~II gas at even smaller radii (around $10^{17}$ cm) from the SN.
However, given the small range of wavelengths covered by the SED above the ionising energy of Ca~II, 
the exclusion range could shift.
Nevertheless, the closely matching profiles of Ca~II~H and Na~I~D indicate that the two gases have the same distribution in the ISM.

Assuming that Na~I and Ca~II trace CS dust, the nondetections of varying absorption lines
exclude a large range of radii from the explosion where dust could have been situated.
Importantly, the earliest spectrum allowed us to probe for dust within $\sim10^{18}$ cm.
Thus, we have been able to test regions where dust can have a significant effect on the light curve 
and observed $R_V$ of a SN \citep{2011ApJ...735...20A,2008ApJ...686L.103G}.

With the two spectra of iPTF~13dge, we have only been able to test a narrow range of radii for photoionisation.
The nondetection of variations in Na~I~D indicate that the gas is situated farther than $10^{18.5}$ cm from the explosion.
Hence, in this case the Na~I gas is also not located in the CS environment.
The intriguing blueshifted feature in the Ca~II~H\&K profiles which is entirely absent in Na~I~D raises 
the speculative possibility that a corresponding Na~I~D feature has already been fully ionised.
If we trust the computed photoionisation curves of Ca~II, this could have occurred in a sweet 
spot around $\sim10^{17.5}$ cm from the explosion.
In the absence of an earlier spectrum where the hypothetical Na~I~D feature could still be present, or a later spectrum 
where the Ca~II~H\&K feature could be partially ionised, this interpretation remains speculative.

We have seen that it is possible to constrain the presence of gas in the CS environment of SNe~Ia
by searching for variations in their absorption profiles due to photoionisation at early phases.
The SED of SNe~Ia indicates that gases such as Na~I, K~I, and Mg~I within a radius of $10^{18}$ cm 
will entirely ionise before SN maximum brightness.
Ca~II would probably ionise slightly later, if at all, within these radii.
In the search for varying absorption lines, spectral time series have so far mostly been taken 
after SN maximum brightness \citep[e.g.,][]{2014MNRAS.443.1849S}.
At these phases, gases in the CS environment will been entirely ionised, reducing the observations to be
sensitive to absorption-line variations caused only by recombination. 
Lastly, SNe with lines of sight that are clear of obvious dust lanes and 
far from their host-galaxy cores 
should be preferred in order to reduce effects originating from ISM gas and dust.
To detect photoionisation effects on CS gases in the future, it is necessary to 
sample isolated SNe~Ia with high-resolution spectra at early phases. 
Specifically, a high-resolution spectrum earlier than $\sim8$ days before maximum is required to 
probe regions within $10^{18}$~cm from SNe using Na~I~D absorption.

\begin{acknowledgements}
We would like to thank Alexis Brandeker for assisting us with the UVES data, 
Jesper Sollerman for his helpful comments, and Daniela Vergani for sharing graphs of the VLA H~I data.
R.A. and A.G. acknowledge support from the Swedish Research Council and the Swedish Space Board. 
The Oskar Klein Centre is funded by the Swedish Research Council.
A.V.F.'s research was funded by US NSF grant AST-1211916,
the TABASGO Foundation, and the Christopher R. Redlich Fund.

This work is based on observations collected at the European Organisation for Astronomical Research in the Southern Hemisphere under ESO programme 091.D-0352(A).
We made use of \swift/\uvot data reduced by P. J.~Brown and released in the {\it Swift} Optical/Ultraviolet Supernova Archive (SOUSA).   
SOUSA is supported by NASA's Astrophysics Data Analysis Program through grant NNX13AF35G.
This work is based on observations made with the Nordic Optical Telescope, operated by the Nordic Optical Telescope Scientific Association at 
the Observatorio del Roque de los Muchachos, La Palma, Spain, of the Instituto de Astrofisica de Canarias.
The data presented here were obtained in part with ALFOSC, which is provided by the Instituto de 
Astrofisica de Andalucia (IAA) under a joint agreement with the University of Copenhagen and NOTSA.
This work makes use of observations from the LCOGT network.
Some of the data presented herein were obtained at the W. M. Keck Observatory, 
which is operated as a scientific partnership among the California Institute of Technology, 
the University of California, and NASA;
the Observatory was made possible by the generous financial support of the W. M. Keck Foundation.
The authors wish to recognise and acknowledge the very significant cultural role 
and reverence that the summit of Mauna Kea has always had within the indigenous Hawaiian community; 
we are most fortunate to have the opportunity to conduct observations from this mountain.
This research used resources of the National Energy Research Scientific Computing Center, 
a DOE Office of Science User Facility supported by the Office of Science of the U.S. Department of Energy under Contract No. DE-AC02-05CH11231.
LANL participation in iPTF was funded by the US Department of Energy as part of the Laboratory Directed Research and Development program.
A portion of this work was carried out at the Jet Propulsion Laboratory, California Institute of Technology, 
under contract with the National Aeronautics and Space Administration.

\end{acknowledgements}

%
%

\bibliographystyle{aa} 
\bibliography{UVES} 

\Online

\begin{appendix} 
\section{Bootstrap}
\label{ap:bootstrap}

   \begin{figure*}
   \centering
       \resizebox{\hsize}{!}{\includegraphics{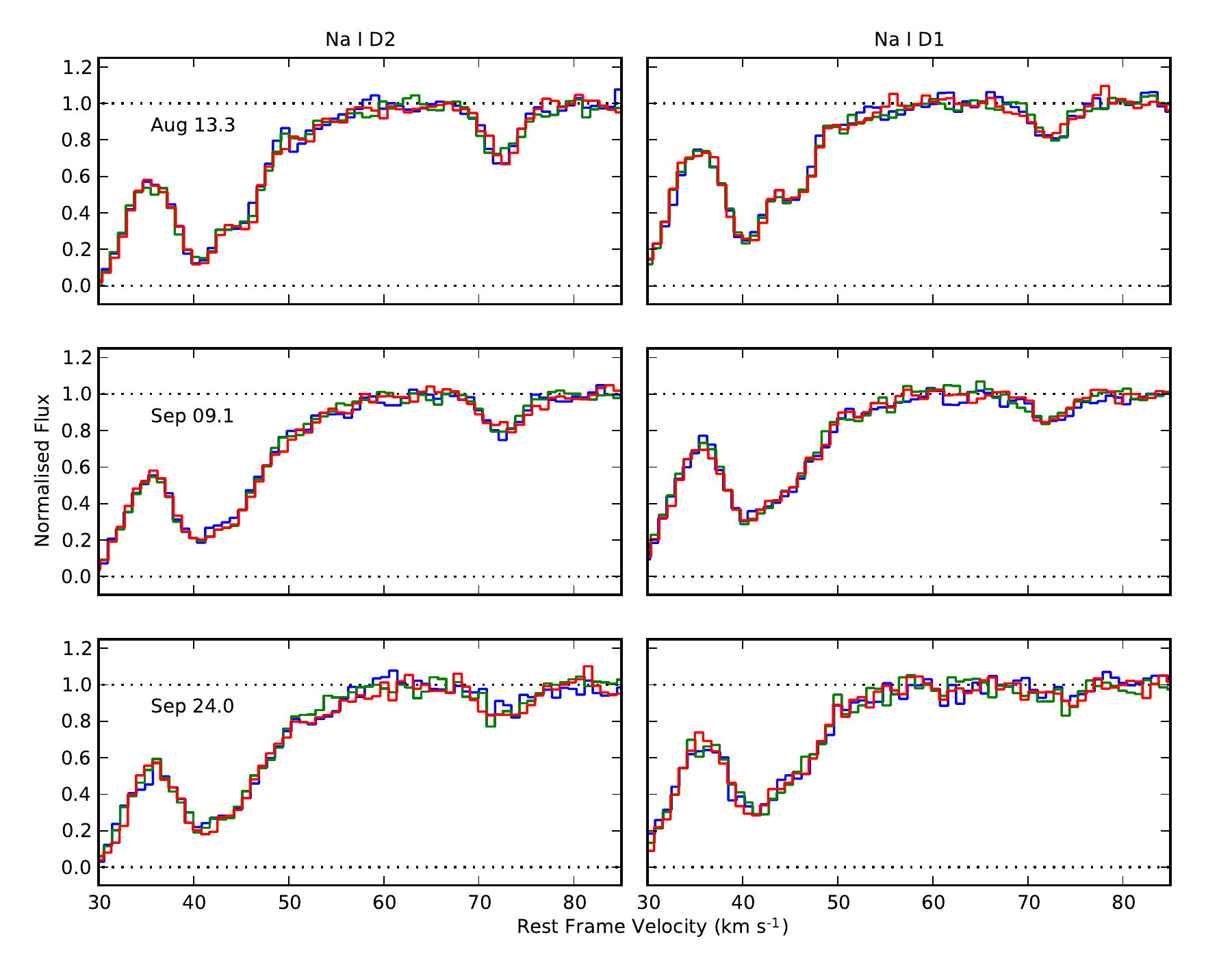}}
       \caption{A comparison of all exposures of the redshifted Na~I~D profile of SN~2013gh.
       Exposures made at the same epochs are in good agreement with each other. }
              \label{fig:all_exp}
    \end{figure*}

\begin{table}
  \centering
  \begin{tabular}{l@{\,}r r@{\,}l r@{\,}l r@{\,}l r@{\,}l}
    \hline\hline
     \multicolumn{2}{c}{\small Epoch}  &  \multicolumn{4}{c}{\small Equivalent Widths} \\ 
     & & \multicolumn{2}{c}{\small Na~I~D1 (m\AA)} & \multicolumn{2}{c}{\small Na~I~D2 (m\AA)}  \\
    \hline
	{\small Aug.} & {\small 13.3} & {\small 16.1} & {\small $\pm$ 2.9} & {\small 30.3} & {\small $\pm$ 2.5} \\
	&& {\small 16.2} & {\small $\pm$ 2.7} & {\small 32.1} & {\small $\pm$ 2.7}\\
	&& {\small 15.7} & {\small $\pm$ 2.7} & {\small 26.1} & {\small $\pm$ 2.2}\\
	\hline
	{\small Sep.} & {\small 9.1} & {\small 20.1} & {\small $\pm$ 2.2} & {\small 23.3} & {\small $\pm$ 2.6} \\
	&& {\small 13.5} & {\small $\pm$ 1.8} & {\small 20.4} & {\small $\pm$ 2.0}\\
	&& {\small 12.9} & {\small $\pm$ 2.3} & {\small 22.7} & {\small $\pm$ 1.9}\\
	\hline
	{\small Sep.} & {\small 24.0} & {\small 3.0} & {\small $\pm$ 3.8} & {\small 18.4} & {\small $\pm$ 4.4} \\
	&& {\small 14.1} & {\small $\pm$ 3.5} & {\small 17.4} & {\small $\pm$ 4.1}\\
	&& {\small 8.8} & {\small $\pm$ 3.1} & {\small 19.8} & {\small $\pm$ 2.9}\\
    \hline\hline
    \multicolumn{1}{l}{}\\
  \end{tabular}
  \caption{The measured EWs of all exposures of the feature at $v\approx73$~km~s$^{-1}$ in SN~2013gh.
    \label{tab:change}}
\end{table}

   \begin{figure}
   \centering
       \resizebox{\hsize}{!}{\includegraphics{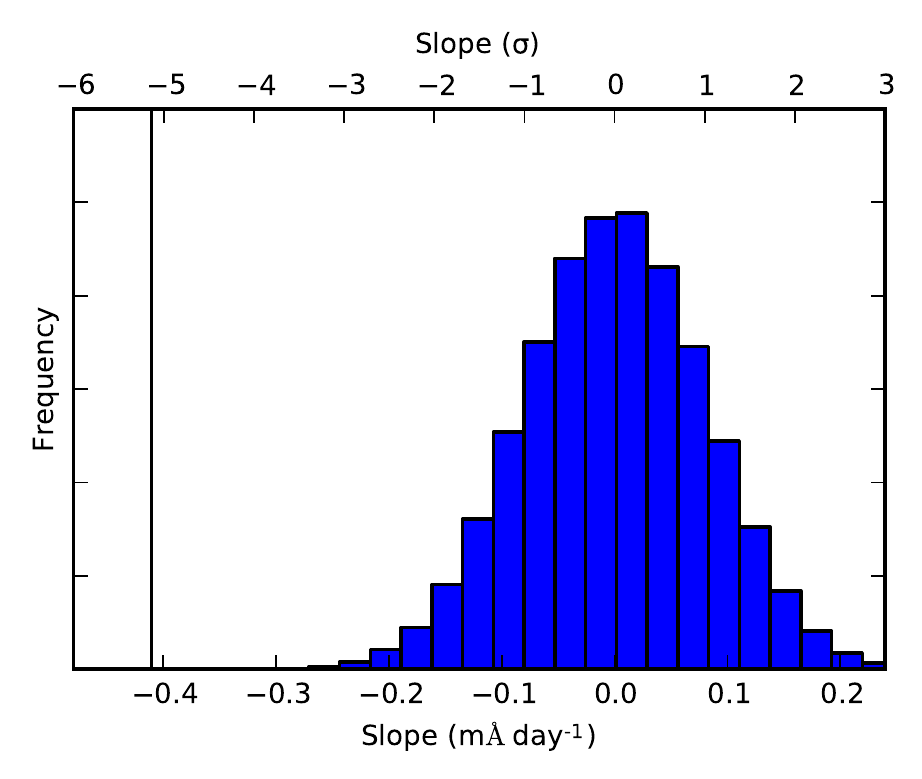}}
       \caption{Distribution of slopes obtained by fitting a linear function to 
       the combined equivalent width (D1$+$D2) values of $10^5$ mock sets of spectra.
       The solid vertical line indicates the slope of the linear fit to the real data.}
              \label{fig:bootstrap}
    \end{figure}

To convince ourselves that the change detected with the Na~I~D Voigt profile fits of SN~2013gh 
is real (see Section~\ref{ssec:change}), one should be able to confirm the result from the EW measurements alone.
First, the absence of significant differences between exposures at the same epochs (see Figure~\ref{fig:all_exp})
should reassure us that the measured changes are real and not caused by spurious fluctuations.
The EWs of the $v\approx73$~km~s$^{-1}$ feature measured from all exposures can be found in Table~\ref{tab:change}.

Since the change in EW of the redshifted Na~I~D feature of SN~2013gh is small,
we would like to ascertain that the decrease is significant.  
We tested the significance with a bootstrap method constraining the likelihood of obtaining the result from random noise.
For this we modelled the feature using the profile parameters obtained from the Voigt profile fit of the weighted averaged 
spectrum of all exposures. To the model absorption profiles, we added normal noise according to the S/N
measured for the respective epochs. 
Next, we measured the resulting EWs of the features in the mock spectra. 
With the null hypothesis that the EW is {\it not} changing between the epochs, we fit
a linear function to the  combined equivalent width (Na~I~D1$+$D2) versus time.
Finally, we determined the distribution of slopes obtained from the mock data to compare with the slope of the real measurements.
Figure~\ref{fig:bootstrap} shows the distribution of slopes of $10^5$ mock spectra. 
The vertical solid line marks the slope of the linear fit to the real observed data, which is $\sim5\sigma$ from the distribution mean. 
This suggest that the variations cannot be the result of random noise.

\end{appendix}

\end{document}

%% file: data_table_short.tex
\begin{tabular}{rrr}
\hline\hline
\multicolumn{1}{c}{MJD} & \multicolumn{1}{c}{Filter} & \multicolumn{1}{c}{Mag}\\
\hline
\multicolumn{3}{c}{SN2013gh}\\
\hline
$56515.1$ & LCOGT r & $15.95(0.06)$\\
$56516.2$ & LCOGT r & $15.57(0.06)$\\
$56517.1$ & LCOGT r & $15.40(0.06)$\\
$56518.1$ & LCOGT r & $14.94(0.06)$\\
$56519.1$ & LCOGT r & $14.76(0.06)$\\
$56520.3$ & LCOGT r & $14.51(0.06)$\\
$56521.1$ & LCOGT r & $14.40(0.06)$\\
$56525.2$ & LCOGT r & $14.00(0.06)$\\
$56526.3$ & LCOGT r & $13.89(0.06)$\\
$56528.7$ & LCOGT r & $13.82(0.06)$\\
$56529.1$ & LCOGT r & $13.85(0.06)$\\
\multicolumn{3}{c}{$\cdots$}\\
\hline
\hline
\multicolumn{3}{c}{}\\
\end{tabular}

%% file: color_table_short.tex
\begin{tabular}{lrlrrrcrrrrr}
\hline\hline
\multicolumn{1}{c}{\small MJD} & \multicolumn{1}{c}{\small Phase} & & \multicolumn{1}{c}{\small\it X} & {\small $A^\mathrm{MW}_\mathrm{\it X}$} & {\small $K_\mathrm{\it X}$} & & \multicolumn{1}{c}{\small {\it V}} & {\small $A^\mathrm{MW}_\mathrm{\it V}$} & {\small $K_\mathrm{\it V}$} & \multicolumn{1}{c}{\small $(\mathrm{\it V}-\mathrm{\it X})_0$}\\
\multicolumn{1}{c}{\small (days)} & \multicolumn{1}{c}{\small (days)} &  \multicolumn{1}{c}{\small Filter} & \multicolumn{1}{c}{\small (mag)} & {\small (mag)} & {\small (mag)} &  {\small Match} & \multicolumn{1}{c}{\small (mag)} & {\small (mag)} & {\small (mag)} & \multicolumn{1}{c}{\small (mag)}\\
\hline
\multicolumn{11}{c}{SN2013gh}\\
\hline
{\small $56518.1$} & {\small $-9.4$} & LCOGT r & {\small $14.94(0.06)$} & {\small $0.06$} & {\small $-0.01$} & M & {\small $15.26(0.09)$} & {\small 0.08} & {\small -0.01} & {\small\ \ \ $0.15$\ \ \ }\\
{\small $56519.1$} & {\small $-8.4$} & LCOGT r & {\small $14.76(0.06)$} & {\small $0.07$} & {\small $-0.01$} & M & {\small $15.04(0.09)$} & {\small 0.08} & {\small -0.01} & {\small\ \ \ $0.13$\ \ \ }\\
{\small $56520.3$} & {\small $-7.0$} & LCOGT r & {\small $14.51(0.06)$} & {\small $0.07$} & {\small $-0.01$} & M & {\small $14.84(0.09)$} & {\small 0.08} & {\small -0.01} & {\small\ \ \ $0.12$\ \ \ }\\
{\small $56521.1$} & {\small $-6.2$} & LCOGT r & {\small $14.40(0.06)$} & {\small $0.07$} & {\small $-0.01$} & D & {\small $14.65(0.07)$} & {\small 0.08} & {\small -0.01} & {\small\ \ \ $0.11$\ \ \ }\\
{\small $56525.2$} & {\small $-1.9$} & LCOGT r & {\small $14.00(0.06)$} & {\small $0.07$} & {\small $-0.01$} & M & {\small $14.22(0.09)$} & {\small 0.08} & {\small -0.01} & {\small\ \ \ $0.11$\ \ \ }\\
{\small $56526.3$} & {\small $-0.7$} & LCOGT r & {\small $13.89(0.06)$} & {\small $0.07$} & {\small $-0.01$} & M & {\small $14.16(0.09)$} & {\small 0.08} & {\small -0.01} & {\small\ \ \ $0.11$\ \ \ }\\
{\small $56528.7$} & {\small $1.9$} & LCOGT r & {\small $13.82(0.06)$} & {\small $0.07$} & {\small $-0.01$} & D & {\small $14.00(0.07)$} & {\small 0.08} & {\small -0.01} & {\small\ \ \ $0.10$\ \ \ }\\
{\small $56529.1$} & {\small $2.4$} & LCOGT r & {\small $13.85(0.06)$} & {\small $0.07$} & {\small $-0.01$} & M & {\small $14.06(0.09)$} & {\small 0.08} & {\small -0.01} & {\small\ \ \ $0.10$\ \ \ }\\
{\small $56530.2$} & {\small $3.5$} & LCOGT r & {\small $13.87(0.06)$} & {\small $0.07$} & {\small $-0.01$} & M & {\small $14.05(0.09)$} & {\small 0.08} & {\small -0.01} & {\small\ \ \ $0.09$\ \ \ }\\
{\small $56531.2$} & {\small $4.6$} & LCOGT r & {\small $13.85(0.06)$} & {\small $0.07$} & {\small $-0.01$} & M & {\small $14.05(0.09)$} & {\small 0.08} & {\small -0.01} & {\small\ \ \ $0.07$\ \ \ }\\
{\small $56532.2$} & {\small $5.7$} & LCOGT r & {\small $13.91(0.06)$} & {\small $0.07$} & {\small $-0.01$} & M & {\small $14.06(0.09)$} & {\small 0.08} & {\small -0.01} & {\small\ \ \ $0.06$\ \ \ }\\
\multicolumn{11}{c}{\small $\cdots$}\\
\hline
\hline
\multicolumn{11}{c}{}\\
\end{tabular}